\documentclass[aapm,graphicx,article,superscriptaddress,amsmath,amssymb]{revtex4-1}

\usepackage[pdftex]{graphicx}
\usepackage{epstopdf}
\usepackage{amsthm}
\usepackage{color}
\usepackage{dcolumn}
\usepackage{tikz}
\usetikzlibrary{arrows}
\usepackage{multirow}
\usepackage{array}
\usepackage{subfigure}
\DeclareGraphicsExtensions{.eps}
\usepackage[mathlines]{lineno}

\newtheorem{prop}{Proposition}

\begin{document}

\preprint{AAPM/123-QED}

\title[Boeck et al.]{On the Interplay between Robustness and Dynamic Planning for Adaptive Radiation Therapy} 
\thanks{There is no COI or financial disclosure to report.}

%

\author{Michelle B\"ock}
\email{miboeck@kth.se}
\affiliation{KTH Royal Institute of Technology, Stockholm, Sweden}
\affiliation{RaySearch Laboratories AB, Stockholm, Sweden}
\author{Kjell Eriksson}
\affiliation{RaySearch Laboratories AB, Stockholm, Sweden}
\author{Anders Forsgren}
\affiliation{KTH Royal Institute of Technology, Stockholm, Sweden}

\date{\today}

\begin{abstract}
Interfractional geometric uncertainties can lead to deviations of the actual delivered dose from the prescribed dose distribution.
To better handle these uncertainties during the course of treatment, the authors propose a dynamic framework for robust adaptive radiation therapy in which a variety of robust adaptive treatment strategies are introduced and evaluated.

This variety is a result of optimization variables with various degrees of freedom  within robust optimization models that vary in their grade of conservativeness.
The different degrees of freedom in the optimization variables are expressed through either time-and-uncertainty-scenario-independence, time-dependence or time-and-uncertainty-scenario-dependence, while the robust models are either based on  expected value-, worst-case- or conditional value-at-risk-optimization.                                                                                                                                                                                                                                                                                                                                                                                                                                                                                                                                                                                                                                                                                                                                                                                                                                                                                                                                                                                                                                                                                                                                                                                                                                                                                                                                                                                                                                                                                                                                                                                                                                                                                                                                                                                                                                                                                                                                                                                                                                                                                                                                                                                                                                                                                                                                                                                                                                                                                                                                                                                                                                                                                                                                                                                                                                                                                                                                                                                                                                                                                                                                                                                                                                                                                                                                                                                                                                                                                                                              The goal of this study is to understand which mathematical properties of the proposed robust adaptive strategies are relevant such that the accumulated dose can be steered as close as possible to the prescribed dose as the treatment progresses. 

We apply a result from convex analysis to show that the robust non-adaptive approach under conditions of convexity and permutation-invariance is at least as good as the time-dependent robust adaptive approach, which implies that the time-dependent problem can be solved by dynamically solving the corresponding time-independent problem.
According to the computational study, non-adaptive robust strategies may provide sufficient target coverage comparable to robust adaptive strategies if the occurring uncertainties follow the same distribution as those included in the robust model.
Moreover, the results indicate that time-and-uncertainty-scenario-dependent optimization variables are most compatible with worst-case-optimization, while time-and-uncertainty-scenario-independent find their best match with expected value optimization.  

In conclusion, the authors introduced a novel framework for robust adaptive radiation therapy and identified mathematical requirements to further develop robust adaptive strategies in order to improve treatment outcome in the presence of interfractional uncertainties.

\end{abstract}
%

\maketitle                    


\section{Introduction}\label{sec:intro}
Adaptive radiation therapy~(ART) is a treatment approach where the treatment plan can be modified in response to patient-specific interfractional geometric variations that could not have been accounted for during the planning process. 
Imaging of the daily patient anatomy and analysis of the impact of anatomical variations on the dose is crucial for adaptive replanning. 
Since image guided radiation therapy~(IGRT) has been integrated in daily clinical routine, interfractional variations can be monitored throughout the course of treatment. 

Moreover, the inclusion of patient-specific variations, which are monitored during the course of treatment, gives the opportunity to provide individualized treatment for each patient. 
In recent years, adaptive radiation therapy has been the subject of a number of clinical and theoretical studies, which demonstrate that this technique can improve treatment outcome~\cite{Lof1998,DeLaZerda2007,Lu2008,Arcangeli2002,Kim2012,Kim2015}.
While theoretical studies focus on the problem formulation and choice of optimization variables, clinical studies improve our understanding of the impact of various factors on the treatment outcome. 
The majority of clinical studies have been conducted on a sample of in-house treated patients with the objective to study the potential benefits of adaptation through re-optimization~\cite{schwartz2013adaptive,Li2013,olteanu2014comparative} or a plan-of-the-day approach~\cite{Qi2014,Hafeez2017,Back2017}, and how to make the resulting workload manageable in daily practice. 
As a consequence, these clinical studies provide valuable insights on clinical aspects such as realistic adaptation frequency during the course of treatment and meaningful criteria that trigger adaptation in the event of non-negligible variations.  

However, the treatment plan should also be robust to geometric interfractional uncertainties, such that variations in the delivered dose distribution are small. 
In conventional treatment planning, interfractional variations are handled by applying safety margins around the target and organs-at-risk~(OAR) in order to generate plans that are robust to errors. Usually, safety margins are obtained from population-based recipes based on systematic and random uncertainties, such as the formulas proposed by van Herk et al.~\cite{VanHerk2000}. Adding these margins around the clinical target volume~(CTV) in order to create the planning target volume~(PTV) thereby increases the treated volume and may consequently increase organ- and normal tissue toxicity while controlling the tumor dose.

As alternatives to safety margins, various frameworks exploiting robust optimization have been proposed~\cite{Unkelbach2009,Unkelbach2004,Albin2012,AlbinRasmus2016} which take into account the possible dose distributions of \textit{a priori} error scenarios during the optimization process.  
These frameworks are based on either expected value optimization~\cite{Unkelbach2004,Unkelbach2009}, i.e., minimizing the expected value of the objective function over the \textit{a priori} scenario doses, or worst-case-optimization~\cite{Albin2012,AlbinRasmus2016}, i.e., minimizing the value of the objective function of the worst-case scenario. 
Bokrantz and Fredriksson~\cite{AlbinRasmus2016} introduced a scenario-based generalised optimization approach to planning with safety margins,
which implicitly generates a PTV-like volume around the target.
 
According to these studies~\cite{Unkelbach2009,Unkelbach2004,Albin2012,AlbinRasmus2016}, robust target coverage and low OAR doses can be achieved by incorporating probabilities of error scenarios into the optimization process. 
Therefore, combining ART with robust optimization may allow for maintaining high treatment quality by responding to interfractional geometric variations which could not have been predicted during the planning process. 
This assumption has been confirmed in our previously published theoretical study~\cite{Boeck2017} on robust adaptive radiation therapy strategies, which suggests that adapting an initial robust plan eventually improves the final therapeutic outcome in comparison to the non-adaptive conventional approach. 
Still, further improvements may be possible through extending the framework to a dynamic adaptive framework. 

One of the first adaptive frameworks based on concepts of control theory was introduced by L\"{o}f et al.~\cite{Lof1998}, which aims at maximizing the tumor control probability by computing the optimal corrections to geometric setup errors and random variations in the beam profiles.  
The issue of dealing with geometric uncertainties and their impact on the fraction dose has been addressed by De La Zerda et al.~\cite{DeLaZerda2007} who propose a closed-loop control framework for adaptive radiation therapy, in which radiation plans are reoptimized in response to change in geometry and thus the delivered dose.  
Another approach to adaptive radiation therapy, introduced by Lu et al.~\cite{Lu2008}, is designed to adapt the fraction dose in response to the varying distance between tumor and critical structure. 
In this particular study, the adapted plans are generated by solving a dynamic programming problem for each fraction. 
However, none of these recently discussed frameworks employ time-and-scenario-dependent optimization variables, which gives the largest number of degrees of freedom in the optimization variables. 
Moreover, the majority of adaptive frameworks in the literature address the issue of finding the optimal fractionation schedule based on biological models for radiation response~\cite{Arcangeli2002,Kim2012,Kim2015,Saberian2016}. 
These publications aim at personalizing the treatment schedule and fraction size on the basis of the modeled dose response of the tumor and healthy tissue. 
These proposed methods are based on stochastic- and optimal control as proposed by Arcangeli et al.~\cite{Arcangeli2002}, Kim et al.~\cite{Kim2012,Kim2015} and Saberian et al.~\cite{Saberian2016} in which adaptation is carried out in response to modeled interfractional variations in the number of tumor cells or varying levels of hypoxia, respectively. 
However, radiobiological models are computationally expensive and considered  simplistic, and therefore hardly used in clinical practice~\cite{Bentzen2010,AllenLi2012}.

In this paper, we present a novel dynamic planning framework for robust adaptive radiation therapy in order to handle interfractional geometric variations, which are included in the robust models as discretized uncertainty scenarios. 
The novelty of our framework is the dynamic approach to robust planning for adaptive radiation therapy which uses time-and/or-scenario-dependent optimization variables combined with various robust models to design a feedback-control framework such that the robust adaptive plan for the subsequent fraction is optimized after every fraction in response to the actual delivered dose. 
As part of the proposed framework, we evaluate a variety of robust adaptive strategies in order to study the relationship between robustness and adaptive replanning.
This variety of robust adaptive strategies is a consequence of choosing different grades of conservativeness and degrees of freedom of the optimization variables. 
Concerning the grade of conservativeness, the strategies are based on either expected value-,  worst-case-, or conditional-value-at-risk~(CVaR) optimization.
The variety in degrees of freedom of the optimization variables stem from using either time-and-scenario-dependent, time-dependent or time-and-scenario-independent optimization variables in the robust adaptive strategies.
In particular, we want to study the resulting treatment quality of those strategies which use time-and-scenario-dependent optimization variables in order to evaluate potential gains from optimizing with respect to predictions on realized uncertainty scenarios over the whole treatment horizon or shorter horizons.
The latter approach is inspired by model-predictive control~(MPC) which is widely used in signal-processing and chemical engineering~\cite{Richalet1993,Morari1999}, but its characteristic look-ahead feature has not yet been exploited in adaptive radiation therapy to the best of our knowledge. 
Specifically, we utilize this characteristic feature in order to (i)~provide predictions of future accumulated dose over a given finite time horizon and (ii)~to compute the optimal adaptive plan required to steer the predicted system output as close as possible to the pre-set target value. 
Since the use of time-and/or-scenario-dependent adaptive plans, i.e. optimization variables, provides more degrees of freedom compared to conventional non-adaptive planning, the adaptive plans are expected to perform better in the presence of interfractional uncertainties. 

The contribution of the mathematical study is the comparative analysis of the mathematical properties of the various robust adaptive and non-adaptive robust strategies, which provides valuable insights into the potential gains of using time-and/or-scenario dependent optimization variables over conventional non-adaptive planning.
As a consequence, we develop a better understanding of the mathematical requirements for robust adaptive strategies such that significant improvements compared to non-adaptive robust planning can be expected. 

To the best of our knowledge, this work represents the first approach of its kind to address the mathematical foundation of dynamic planning for robust adaptive radiation therapy, which introduces a proof-of-concept framework to analyse various robust adaptive strategies.
The framework focuses on handling geometric variations only and at this point does not take into account radio-biological response. Biologically-based models are hardly used clinicial routine and should be used with caution~\cite{Bentzen2010,AllenLi2012}. 
In order to focus on the mathematical concepts of the presented framework and reduce computational effort a simplified one-dimensional model is used. 
However, this simplified representation contains relevant aspects of robustness, dynamic planning and treatment plan adaptation providing valuable insights, which is crucial before applying such a framework to clinical patient data.  

\section{Methods}
Usually, patients experience interfractional geometric variations throughout the whole course of treatment.
In our proposed framework, interfractional uncertainties are handled by a variety of robust adaptive strategies. 
Moreover, we assume to have complete information on the interfractional uncertainties and their impact on the delivered dose.
\subsection{Notation}
Throughout this work~$T$ denotes the total number of fractions~$t$, while~$N$ denotes the total number of voxels~$n$.
The accumulated dose at fraction~$t$ is denoted by~$x_t \in \mathbb{R}^N$ which in our framework is referred to as the system at the sampling instant~$t$.
Since it is assumed that the patient does not receive any radiation before the start of the treatment at~$t=0$, the dose~$x_0$ is set to zero. 
The radiation plan to be applied at~$t+1$ is given by~$u_t\in \mathbb{R}^M$, where~$M$ refers to the number of bixels. In our framework the plans are referred to as control signals.
Dose deposition is modeled by a dose deposition matrix~$B\in \mathbb{R}^{N \times M}$. 
In a deterministic setting, the accumulated dose would be given by~$x_{t+1}= x_t + Bu_t$ for~$t=0,1,\dots,T-1$. 
However, the focus of our framework is on handling interfractional uncertainties. 
In this work, the interfractional variations~$\omega$ are assumed to vary from fraction to fraction and thus, modeled as independent and identically distributed (i.i.d.)~random variables.
The uncertainties are derived from a discretized normal distribution~$\mathcal{N}(0,\sigma^2)$ and summarized in the set~$\Omega$.
The probability of each i.i.d.~uncertainty scenario~$\omega \in \Omega$ to occur is denoted by~$p_{\omega}$.
The geometric uncertainty occurring at fraction~$t$ is modeled by the stochastic shift matrix~$S(\omega_t)\in \mathbb{R}^{M \times M}$.
In our framework, the impact of interfractional uncertainties on the fractionated treatment is modeled by the linear, stochastic, discrete-time state-space dynamics
\begin{equation}\label{eq:stochSystDyn}
x_{t+1} = x_t + BS(\omega_{t+1})u_t \quad \text{for} \ t = 0,1,\dots,T-1.			
\end{equation}								   		

\subsection{Robust Adaptive Strategies}\label{subsec:strat}
Our proposed framework for robust adaptive radiation therapy uses dynamic optimization variables in order to deliver the prescribed dose distribution~$d_T$ in the presence of uncertainties. 
In this study, the goal is to achieve a uniform dose in CTV and zero elsewhere.
In order to minimize the gap between the accumulated dose~$x_T$ and the prescribed dose~$d_T$, we choose the quadratic penalty~$\left\Vert  x_T -d_T \right\Vert^2$, i.e.,~$f \left(  \left\Vert  x_T -d_T \right\Vert^2 \right)$ as a planning objective which is widely used in clinical planning systems, where~$f$ refers to the particular robust optimization model.
A variety of robust adaptive strategies are introduced with the purpose to better understand how to handle interfractional uncertainties.
It is the goal of every robust adaptive strategy to generate an individual plan~$u_t$ at fraction~$t$ which is going to be delivered at fraction~$t+1$.
The robust adaptive strategies are categorized based on conservativeness, the degrees of freedom of their optimization variables and their timing for treatment planning.
In order to clearly distinguish between the time instant of plan delivery~$t$ and planning, we introduce the index~$t_P$ for the latter. 
The optimization variables~$u_t$, i.e., plans are either i) time-and-scenario-independent, ii) time-dependent or iii) time-and-scenario-dependent.
Time- and scenario-independent optimization variables are used in the \textit{non-adaptive robust strategy}, which refers to applying the same plan throughout the course of treatment.
Time-dependent optimization variables are the basis for the \textit{time-varying robust} and \textit{time-varying robust adaptive strategy}.
Time- and scenario-dependent optimization variables are used in the \textit{time- and scenario-dependent robust adaptive strategy} and the robust adaptive strategies based on model-predictive control~(MPC), referred to as the \textit{MPC-strategies}.
The use of time- and scenario-dependent optimization variables provides the most degrees of freedom to mitigate the effect of interfractional uncertainties on the accumulated dose, while time- and scenario-independent optimization variables provide the least. 
The performance of the robust adaptive strategies is analysed in comparison with the corresponding non-adaptive robust strategy. 

As previously discussed, it is the overall goal of all robust strategies introduced in this framework to minimize the quadratic penalty of the gap between the final accumulated dose~$x_T = x_0 + \sum_{\tau=0}^{T-1}BS(\omega_{\tau+1})u_{\tau}$ and the prescription dose~$d_T$, but they all differ in their approach to generate the plans~$u_t$ to achieve that goal. 
Ultimately, it is the objective of our study to analyse the mathematical properties and performance of the proposed robust adaptive strategies to solve
\begin{equation}\label{eq:TV-Rob}
\begin{aligned}
\underset{u_{\tau}\geq 0}{\text{min}} & \ f\left(  \left\Vert x_{t_p} + \sum_{{\tau}=t_P}^{T-1}BS(\omega_{\tau+1})u_{\tau}  - d_T   \right\Vert^2 \right), \\ 
\text{s.t.} & \ x_{\tau+1} = x_{\tau} + BS(\omega_{\tau+1})u_{\tau} \quad \text{for} \ \tau= t_P,t_P+1,\dots,T-1;  \  x_{t_P} \ \text{given}, 
\end{aligned}
\end{equation}
where the current system state~$x_{t_p}$ is assumed to be always known and~$f$ refers to the specific robust model that is combined with a specific treatment strategy.
In particular, we are interested in analysing the performance of combining expected value-, worst-case- or conditional-value-at-risk~(CVaR)-optimization with non-adaptive and adaptive strategies.

\subsubsection{Non-adaptive robust strategy}
The \textit{non-adaptive robust strategy} is introduced and evaluated in our proof-of-concept framework in order to represent the conventional non-adaptive approach to radiation therapy in which the same plan is delivered at every fraction throughout the course of treatment.
Since it is assumed that information on the accumulated dose~$x_t$ that could be acquired during the course of treatment is not included during planning, problem~(\ref{eq:TV-Rob}) is solved before the start of the treatment at~$t_P =0$ for positive time- and scenario-independent optimization variables~$u_{\tau}=u \geq 0 \in \mathbb{R}^M \ \forall \tau$.  
The \textit{non-adaptive robust strategy} is introduced and evaluated in our framework in order to serve as a lower benchmark which allows us to quantify the extent of potential improvements of the various proposed robust adaptive strategies which will be introduced later in this work.


\subsubsection{Time-varying robust strategy}
In contrast to the \textit{non-adaptive robust strategy}, the plans in the \textit{time-varying robust strategy} may vary from fraction to fraction and therefore, the degrees of freedom of the optimization variables are increased to include time-dependency, but information on the current accumulated dose is not taken into account either. 
Thus, problem~(\ref{eq:TV-Rob}) is solved before the start of the treatment at~$t_P =0$ for positive time-dependent optimization variables~$u_{\tau}\geq 0\in \mathbb{R}^M \ \forall \tau$.
With the \textit{time-varying robust strategy}, we want to investigate if adding time-dependency as a degree of freedom to the time-and-scenario-independent optimization variables, used in the non-adaptive robust strategy, may give any advantages.
In fact, the move from time- and scenario-independent to time-dependent variables is not necessary, if (i) the final accumulated dose is evaluated in a quadratic cost function, and if (ii) the uncertainties that occur at every fraction are assumed to follow the same distribution. 
Since conditions (i) and (ii) imply that delivering the mean over all time-dependent plans at every fraction will be at least as good delivering the time-dependent plans at the corresponding fraction.
In mathematical terms, the choice of the quadratic cost function in our model gives convexity, while the evaluation of the final accumulated dose affected by i.i.d.~uncertainties represents the permutation-invariance.
The mathematical evaluation of such a case has been analyzed by Parillo~\cite[p.~100]{Parillo}, as reviewed in the following proposition.
\begin{prop}[Parrilo {\cite[p.~100]{Parillo}}] \label{prop1}
Let~$f: \left[ U \right] \rightarrow \mathbb{R} $ be a convex and permutation invariant function of a convex problem, where~$\left[ U \right]$ refers to the T-tuple~$ ( u_0,u_1,\dots,u_{T-1} )$ of vectors~$u_t \in \mathbb{R}^{N}$ for~$t = 0,1, \dots,T-1$ . 
Then, 
\[f \left( \left[ \bar{U} \right] \right) \leq f \left( \left[ U \right] \right), \]
with~$\left[ \bar{U} \right]$ denoting the T-tuple of the mean~$\frac{1}{T} \sum_{t=0}^{T-1} u_t$ over all~$u_t$ for~$t= 0,1,\dots,T-1$.
\end{prop}
For the sake of completeness, the complete proof is given in appendix~\ref{appendix:Proof}.
This result also indicates that under the conditions of convexity and permutation invariance, the conventional approach to treatment planning implicitly takes fractionation into account.
Therefore, evaluation and analysis of the final dose distribution for the \textit{time-varying robust strategy} are omitted. 

\subsubsection{Time-varying robust adaptive strategy}
In the \textit{time-varying robust adaptive strategy}, the treatment is designed to generate an adaptive plan at every fraction~$t_P=t$
in response to the accumulated dose~$x_t$ and with respect to the prescription dose~$d_T$ and the remaining number of fraction~$T-t$.
In contrast to the previously introduced strategies, planning occurs at every fraction and information on the accumulated dose is included in the planning process. 
As a consequence, we solve~(\ref{eq:TV-Rob}) at every fraction~$t$ after observing~$x_t$ for the remaining number of fractions~$T-t$.
Thus, the problem to be solved at~$t_P = 0$ is equivalent to the \textit{time-varying robust} and \textit{non-adaptive robust strategy}.
However, at~$t_P = t, \ \text{for} \ t =1,2,\dots,T-1$, the plans are optimized with respect to a shrinking treatment horizon~$T-t$.
As a consequence of our analysis of the time-varying robust strategy and Proposition~\ref{prop1}, it is sufficient to apply the \textit{time-varying robust adaptive strategy} with time-and-scenario-independent optimization variables. 
In other words, we solve a problem at every fraction with the same properties as in the \textit{non-adaptive robust strategy}, but with respect to a shrinking treatment horizon~$T-t$ and accumulated dose~$x_t$. 

\subsubsection{Time-and-scenario-dependent robust adaptive strategy}
In the \textit{time-and-scenario-dependent robust adaptive strategy}, plans for every fraction~$t$ and possible realization of uncertainty scenarios~$\tilde{\omega}_1^T =\{\omega_1,\omega_2, \dots ,\omega_T \}$ over the whole course of treatment are generated.
In other words, at every fraction~$t$ an adapted plan is generated by solving~(\ref{eq:TV-Rob}) at~$t_P = t$ in response to the accumulated dose~$x_t$ and the possible uncertainty scenarios to occur in the subsequent fractions~$\tilde{\omega}_{t+1}^T = \{\omega_{t+1},\omega_{t+2}, \dots ,\omega_T \}$.
However, in order to optimize the plan~$u_0$ for the first fraction problem~(\ref{eq:TV-Rob}) is solved at~$t_P = 0$ by taking into account~$\sum_{t=1}^{T}|\Omega|^{t}$ many possible realizations of uncertainties contained in~$\tilde{\Omega}_1^T = \{ \Omega \times \Omega \times \cdots \times  \Omega \}$, where~$\lvert \Omega \rvert$ denotes the cardinality of~$\Omega$. 
Thus, the size of the scenario tree grows exponentially with the number of fractions.
In contrast to the \textit{time-varying robust adaptive strategy}, the plans are created to be time-and-scenario-dependent.
Therefore, the \textit{time-and-scenario-dependent robust adaptive strategy} provides the most degrees of freedom, which comes at the price of taking into account~$\sum_{t=1}^{T}|\Omega|^{t}$ many realizations during the optimization of $\sum_{t=0}^{T-1}|\Omega|^{t}$ many plans at~$t_P=0$ which are going to be applied in the corresponding scenario tree.
As a consequence of using time-and-scenario-dependent variables,  obtaining the optimal set of plans for every possible sequence of uncertainty scenarios may be too computationally expensive and not even viable for a regular number of fractions. 
However, this strategy could be applied to hypofractionated radiation therapy in which the total dose is administered in fewer fractions and therefore higher fraction doses and which may benefit to a great extent from adaptive strategies~\cite{Unkelbach2014,Cantin2015,Fast2016}. 
In general, a compromise between optimality and computational affordability has to be made. 

\subsubsection{MPC-strategies}
In order to achieve a trade-off between optimality and computational affordability while using time-and-scenario-dependent variables, 
we choose to optimize with respect to a relative to~$T$ much shorter planning horizon~$K$. 
We therefore introduce robust adaptive strategies which are based  on model-predictive control~(MPC) in order to take advantage of its look- ahead-feature in treatment planning.
In the MPC-strategies, the adaptive plan is generated by solving a modified version of~(\ref{eq:TV-Rob}) at~$t_P=t$ in which~$T$ and~$d_T$ are replaced by~$\text{min}\{ T, t_P+K \}$ and~$\frac{\text{min}\{ T, t_P+K \}}{T}d_T$, respectively. 
Adaptive plan are optimized throughout the course of the treatment for~$ t = 0,1,\dots T-K-1$ in response to the accumulated dose~$x_t$ and predictions of the accumulated dose~$K$ fractions ahead.
Thus, the MPC-strategies require the computation of at most~$\sum_{j=1}^{K}|\Omega|^j$ dose predictions and~$\sum_{j=1}^{K-1}|\Omega|^{j-1}$ plans, which is lower than in the \textit{time-and-scenario-dependent robust adaptive strategy} since~$K \ll T$.
Therefore, we introduce the \textit{MPC-strategies} as a clinically and computationally feasible compromise. 
In contrast to the previously discussed strategies, we choose to optimize the adaptive plans with the aim to achieve the modified prescription dose~$\frac{t_P+K}{T}d_T$ considering the prediction horizon~$K$.
Nevertheless, the adaptive plans could also be generated with respect to the remaining dose~$(d_T -x_{t_P})$ over the remaining number of fractions~$T-t$ multiplied by the length of the shorter horizon~$K$.
In this case,~$T$ and~$x_{t_P}-d_T$ in~(\ref{eq:TV-Rob}) are replaced by~$\text{min}\{ T, t_P+K \}$ and~$K\frac{(d_T -x_{t_P})}{T-t_P}$, respectively.  
The evaluation of this modified objective function suggested though, that this modified approach may not be superior over our proposed \textit{MPC-strategies}.
Thus, further evaluation of the modified approach is omitted.


\subsection{Robust Models}
In section~\ref{subsec:strat}, we introduced various strategies to solve~(\ref{eq:TV-Rob}) in which~$f$ refers to a specific robust model. 
In this section, we will discuss the nature of the robust models chosen to be combined with the non-adaptive and adaptive strategies.
Expected value optimization is the least conservative model, while worst-case-optimization the most conservative; and conditional-value-at-risk~(CVaR)-optimization represents a compromise between the two. 
Thus,~$f\left( \left\Vert  x_T -d_T \right\Vert^2  \right) $ in~(\ref{eq:TV-Rob}) takes the following form in

\begin{enumerate}
\item Expected-value-optimization 
\begin{equation}\label{eq:Dquadcost}
\mathbb{E} \left[ \left\Vert  x_T -d_T \right\Vert^2_D \right],
\end{equation} 
\item Worst-case-optimization
\begin{equation}\label{eq:Dminimaxcost}
\underset{ \tilde{\omega}_1^T \in \tilde{\Omega}_1^T }{\text{max}} \left\Vert  x_T -d_T \right\Vert^2_D 
\end{equation} and
\item Conditional-value-at-risk~(CVaR)-optimization
\begin{equation}\label{eq:DCVaRcost} \underset{\lambda}{\text{min}} \ \lambda + \frac{1}{\alpha} 
\mathbb{E} \left[ \left(\left\Vert  x_T - d_T \right\Vert^2_D -\lambda \right)_{+} \right],
\end{equation}
\end{enumerate}
where the shorthand~$z_+$ and~$\lambda$ denote~$\max\left\lbrace z, 0 \right\rbrace$ and an additional optimization variable, respectively. 
The matrix~$D$ in~(\ref{eq:Dquadcost}),~(\ref{eq:Dminimaxcost}) and~(\ref{eq:DCVaRcost}) is referred to as the weighting matrix which will be described in more detail later.
In order to avoid discontinuities in the worst-case- and CVaR- optimization models, the objective functions in~(\ref{eq:Dminimaxcost}) and~(\ref{eq:DCVaRcost}) are reformulated as
\begin{equation}\label{eq:LinMinimax}
\underset{z \geq 0}{\text{min}} \ \{ z: \ z \geq  \left\Vert x_T - d_T \right\Vert_D^2  \ \forall \ \tilde{\omega}_1^T \in \tilde{\Omega}_1^T  \}
\end{equation}
and
\begin{equation}\label{eq:dualCVaR}
\underset{\lambda,\zeta}{\text{min}} \ \{ \lambda + \frac{1}{\alpha} p_{\tilde{\omega}}^T \zeta: \ \lambda + \zeta_{\tilde{\omega}}  \geq \left\Vert  x_T-d_T  \right\Vert^2_D \ \forall \tilde{\omega}_1^T \in \tilde{\Omega}_1^T, \ \zeta \geq 0 \},
\end{equation}
the max function is substituted, the help variable~$\zeta$ is introduced and where~$p_{\tilde{\omega}} = p_{\omega} \times p_{\omega} \times \cdots \times p_{\omega}$ refers to the probability vector for the realizations~$\tilde{\omega}_1^T$.  
Since the number of possible realizations of uncertainty scenarios grows exponentially with the number of fractions~$T$, problem reformulations may be considered.
In case of minimizing the expected value of the quadratic penalty~($\ref{eq:Dquadcost}$), the optimal solution can be computed in an exact manner independent of~$T$ by exploiting the structure of the quadratic penalty, which is demonstrated in detail in Appendix~\ref{appendix:Reform_Expect}. 
In worst-case- and CVaR-optimization, an exact optimal solution can be computed depending on computational power and for a much shorter than conventionally used treatment length~$T$. 
In cases such as~$T=30$, the solution could be optimized based on Monte Carlo simulated treatments subjected to interfractional uncertainties~$\omega \in \Omega$.
CVaR-optimization can be interpreted as a compromise between expected value- and worst-case scenario optimization, since its optimal solution minimizes the expected value of the fraction~$0<\alpha \leq 1$ of the worst scenarios~$\omega \in \Omega$~\cite{Albin2012}. 
Thus, the parameter~$\alpha$ which affects the upper bound of~$p_{\omega}$ controls the grade of conservativeness. 
The closer~$\alpha$ is to~$1$, the less conservative the solution will be, while~$\alpha \leq \min_{\omega \in \Omega} p_{\omega}$ with~$p_{\omega} > 0$ changes~(\ref{eq:DCVaRcost}) to be equivalent to worst-case-optimization in~(\ref{eq:Dminimaxcost}). In this study,~$\alpha$ is set to~$0.4$. 
The concept of CVaR-optimization was introduced by Rockafellar and Uryasev~\cite{rockafellar2000} and it is widely used in finance for risk management.
With reference to robust planning, Fredriksson~\cite{Albin2012} introduced a framework for robust planning using CVaR to linearly adjust the grade of conservativeness.

For the sake of readability, we introduce the Euclidean norm with respect to the weighting matrix~$D$ in~(\ref{eq:Dquadcost}),~(\ref{eq:Dminimaxcost}) and~(\ref{eq:DCVaRcost}) in order to simplify the extended quadratic penalty function
\begin{equation}\label{eq:ModPenalty}
\left[ \sum_{r\in\mathcal{R}} g_r \sum_{n\in\mathcal{N}_r} \eta_{n,r} v_{n,r}(x_{T,n}-d_{T,n})^2 \right],
\end{equation}
which combines the conventionally used importance weights~$g_r$, the relative volumes~$v_{n,r}$ satisfying~$\sum_{n\in \mathcal{N}_r} v_{n,r} = 1$ for each region of interest~(ROI)~$r \in \mathcal{R}$ denoting the set of all regions and the scenario-based voxel weights~$\eta_{n,r}$. These scenario-based voxel weights~$\eta_{n,r}$ are a result of a generalized scenario-based robust optimization approach to account for interfractional uncertainties introduced by Fredriksson and Bokrantz~\cite{AlbinRasmus2016}, which omits the use of CTV-PTV margins.
These voxel weights are derived from modeling an uncertainty scenario~$\xi \in \Xi$ as whole-body shifts~$m(\xi)$ and summarizing the voxels part of the CTV under scenario~$\xi$ in the set~$\mathcal{C}'(\omega)=\left\lbrace  n + m(\xi): n\in \mathcal{C} \right\rbrace$, with $\mathcal{C}$ referring to the voxels contained in the CTV under the scenario of no shift occurring. 
Thus, the scenario based voxel weights are equivalent to the reciprocal number of scenarios~$\xi$ under which voxel~$n$ is contained in the scenario-related set~$\mathcal{C}'(\xi)$ as expressed by
\begin{equation}\label{eq:voxelweight}
\eta_n = \frac{1}{\lvert \left\lbrace\xi \in \Xi: n \in \mathcal{C}'(\xi) \right\rbrace \rvert}.
\end{equation}
The operator~$\lvert \cdot \rvert$ in~(\ref{eq:voxelweight}) refers to the cardinality of the set~$\left\lbrace \xi \in \Xi: n \in \mathcal{C}'(\xi) \right\rbrace$ and the scenario-based voxel weights~$\eta_n$ satisfy the relation~$1/{\lvert \Xi \rvert} \leq \eta_n \leq 1$ for every voxel~$n$.

\subsection{Geometry}
\begin{figure}
\begin{center}
\includegraphics[scale=1]{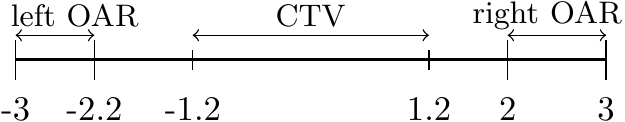}
\caption{One-dimensional patient phantom. The dimensions are given in cm.}\label{fig:Geometry}
\end{center}
\end{figure}

In order to focus on the mathematical properties of the proposed framework, a one-dimensional patient phantom, as illustrated in Figure~\ref{fig:Geometry}, is considered. 
This model schematically represents a slice of a two-dimensional phantom or an intersection of a sagittal and transversal cut of a three-dimensional patient-geometry. The one-dimensional phantom geometry is discretized into 40 voxels. 
The phantom contains one clinical target volume (CTV) and two organs at risk (OARs) which are asymmetrically located around the CTV. 
The CTV is located between -1.2 and 1.2 cm, while the organs at risk are located in the intervals ~$[-2,-2.2]$~cm and~$[2,3]$~cm.
The CTV is an extension of the visible tumor and accounts for the microscopic spread of cancer cells. 
As a consequence, dose coverage of the CTV during the whole course of treatment has to be maintained despite the presence of interfractional variations in order to guarantee a successful treatment outcome.
According to our model, the CTV is irradiated by a perpendicular oriented field, while the absorbed dose in each voxel is modeled by Gaussian functions at a spacing of~1.5~mm with a standard deviation of~3~mm, which is represented by the dose-deposition matrix~$B$. 
As a consequence of the idealized phantom geometry, the number of bixels~$M$ and their positions are assumed to be identical with the number of voxels~$N$ and their location. 
Thus, the dimensions of the plans, the shift-matrix and dose-deposition matrix are simplified to~$u_t \in \mathbb{R}^M$,~$S(\omega)\in \mathbb{R}^{N \times M}$ and ~$B\in \mathbb{R}^{M \times M}$, respectively.
In consideration of reducing computational effort, interfractional geometric uncertainties are modeled as rigid whole-body shifts such that the dose received by each voxel can be computed and tracked in a straightforward manner. 


\subsection{Computational Study}
The computational study is performed in MATLAB version 9.1 using the IBM optimization solver CPLEX in the studio version 12.6.3. 
As previously discussed, we assume that the one-dimensional phantom geometry is subjected to rigid whole-body shifts. 
These random shifts~$\omega \in \Omega$ are derived from the discretized normal distribution~$\mathcal{N}(0,0.25)$, which is in accordance with clinical studies on interfractional prostate motion~\cite{Langen2001,Wu2001,Balter1995}.  
The discretized scenarios~$\omega \in \Omega$ are obtained by adapting the normal distribution to the voxel grid according to its mean and standard deviation~\cite{Boeck2017}. 
The set~$\Omega=\{ -2,-1,0,1,2\}$ which is included in the robust optimization models contains the uncertainty scenarios measured in voxel shifts with the corresponding normalized probabilities~$p_{\omega}=\{ 0.0924, 0.2414, 0.3324, 0.2414, 0.0924 \}$,
Throughout all optimization models and treatment strategies the optimization weight~$g_{CTV}$ for the CTV is set to~$100$, while the weight for the right OAR and left OAR is set to~$10$ and the external is set to~$1$. 
The study of the proposed framework is conducted in two stages.

First, the \textit{feasibility study} is carried out in order to study the trade-off between mathematical optimality and computational feasibility of our proposed framework and the feasibility of adaptive strategies in hypofractionated radiotherapy.
The trade-off is evaluated by comparing the objective function values at the optimal solution of every strategy. 
Among the presented strategies, the non-adaptive robust strategy is considered the lower benchmark and the time-and-scenario-dependent strategy the upper benchmark, since it takes into account uncertainties over the whole treatment horizon.
In the mathematical study, the impact of the choice of optimization variables, i.e.: time-dependent or time-and-scenario-dependent, on the optimal solution is examined.
For the sake of computational tractability, the total number of fractions~$T$ is set to five and the prediction horizon~$K$ is set to one, two and three fractions.
As a consequence, the evaluation of the \textit{time-varying robust adaptive} and \textit{MPC-strategies} relative to the \textit{robust non-adaptive} and \textit{time-and-scenario-dependent robust adaptive strategy} is carried out in a fair manner.  

Second, in the \textit{conventional fractionation length study} we study the performance of our proposed strategies for a population of 100 treatments, each consisting of~$30$ fractions with interfractional uncertainties. 
These uncertainty scenarios are generated from the same normal distribution~$\mathcal{N}(0,0.25)$ as those accounted for during the optimization process, but truncated in a different manner in order to test our strategies for a larger set of uncertainty scenarios~$\{ -3,-2,-1,0,1,2,3 \}$.
For the sake of a fair comparison, the \textit{robust non-adaptive}, \textit{time-varying robust adaptive} and \textit{MPC-strategies} are evaluated for the same treatment population.
The plans for the \textit{robust non-adaptive} and \textit{time-varying robust adaptive} strategy combined with worst-case- and CVaR-optimization for~$T=30$ are computed based on Monte Carlo simulations, as explained previously.
In order to investigate the statistical accuracy and confirm the predictive value of our study, the framework is evaluated for two additional populations with~$150$ and~$200$ treatments. 
From all three populations, the total sum of squares of voxel dose deviations from the prescribed dose and root mean square errors are obtained and compared with each other.
Since these values are of the same order of magnitude, we conclude that the population of 100 treatment is sufficiently large.

\section{Results}
\subsection{Feasibility Study}
In the \textit{feasibility study}, the objective function value for every robust strategy is evaluated at its respective optimal solution and summarized in table~\ref{tab:FeasStudyResult} for a mathematical and objective analysis of the proposed strategies. 
We quantify to which extent the objective function values of the various robust adaptive strategies decrease in comparison with the corresponding non-adaptive robust strategy as the degrees of freedom in the optimization variables increase.
Thus, the comparison is performed among the strategies and not among the robust models.
Moreover, we investigate how well the \textit{MPC-strategies} approximate the \textit{time-and-scenario-dependent strategy}.
The objective function values of the \textit{time-and-scenario-dependent strategy} in combination with CVaR-optimization could not be computed because of computational issues related to limited memory, which underlines the importance of approximation methods, e.g. MPC. 
Overall, the evaluation of objective function values indicates a number of trends. 

First, the objective function values of the robust adaptive strategies are lower than those of the respective non-adaptive strategy, which confirms the property of the \textit{non-adaptive robust strategy} as the lower benchmark and the \textit{time-and-scenario-dependent strategy} as the upper benchmark in the context of mathematical accuracy.
Second, the adaptive strategies which consider uncertainties over a prediction horizon larger than one fraction, i.e., the \textit{time-varying robust adaptive}, \textit{MPC2-} and \textit{MPC3-strategies} achieve lower objective function values than the \textit{MPC1-strategy}, which indicates a likely benefit of taking into account possible uncertainties over a prediction horizon~$K>1$. 
The decrease of objective function value in the \textit{time-and-scenario-dependent} and \textit{MPC-strategies} stems from optimizing the plans~$u_0$ and~$u_t$ with respect to possible realizations of uncertainty scenarios over the horizon~$T$ or~$K$.
However, there is a risk of overcompensation relative to the prescribed fraction dose~$\frac{d_T}{T}$. 
Therefore, we introduce stabilizing constraints to keep the fraction dose within the interval~$(1-\gamma)\frac{d_T}{T} \leq Bu_{\tau} \leq (1+\gamma)\frac{d_T}{T}$, where~$\gamma \geq 0$ specifies the user-defined lower and upper thresholds for~$u_t$. 
In our framework,~$\gamma$ is set to~$0.05$.
The nature of overcompensation is discussed later in this section. 
The trend of decreasing objective function values continues for the stabilized \textit{time-and-scenario-dependent-} and \textit{MPC-strategies} as illustrated in table~\ref{tab:FeasStudyResult}, even though it is less pronounced since the stabilizing constraints decrease the feasible region.
However, in case of combining CVaR optimization with the \textit{MPC1-strategy} this trend may not be as strong as in the other robust models, since its objective function value is slightly larger than that of the \textit{non-adaptive robust strategy}. 
Third, the combination of worst-case-optimization with the \textit{time-and-scenario-dependent-} and \textit{MPC-strategy} seems to lead to the largest improvement in objective function value from the non-adaptive strategy. 
Thus, worst-case-optimization may be most compatible with time-and-scenario-dependent variables. 
It should be pointed out that the \textit{MPC-strategies} are approximations of the \textit{time-and-scenario-dependent strategy} and that the objective function values are computed from all~$|\Omega|^T$ attainable final dose distributions and therefore, small differences in objective function values should be interpreted in relative terms.   
\begin{table}
\center
\caption{Evaluation of the final doses~$x_{5}$ in all models and strategies for $|\Omega| = 5$ and number of voxels $N = 40$. $^{*)}$ The optimal solution of the \textit{time-and-scenario-dependent strategy} combined with CVaR-optimization could not be computed due to computational limitations.}
	\begin{tabular}{|c| >{\centering\arraybackslash}m{2.3cm}|  >{\centering\arraybackslash}m{2.3cm}| >{\centering\arraybackslash}m{1.3cm}| >{\centering\arraybackslash}m{1.3cm}| >{\centering\arraybackslash}m{1.3cm}| >{\centering\arraybackslash}m{3cm}|}
	\hline
& \textit{non-adaptive}&  \textit{time-varying adaptive} & \textit{MPC1}& \textit{MPC2}& \textit{MPC3}& \textit{time- \& scenario dependent} \\ \hline \hline
$\mathbb{E}$&  0.2086  & 0.1868 & 0.1991& 0.1794 & 0.1743 & 0.1729 \\ \hline 
stabilized $\mathbb{E}$& n/a & n/a & 0.2097 & 0.2013 & 0.2003 & 0.2002 \\ \hline \hline
worst-case & 0.4832  & 0.3413 & 0.3431 & 0.3179 & 0.3058& 0.2977 \\ \hline 
stabilized worst-case & n/a & n/a & 0.3479 & 0.3278 & 0.3198& 0.3167 \\ \hline \hline
CVaR  & 0.2552  & 0.2371 & 0.2575 & 0.2302 & 0.2195 & $-^{*)}$ \\ \hline 
stabilized CVaR  & n/a  & n/a & 0.2606 & 0.2445 & 0.2411 & $-^{*)}$ \\ \hline 
	\end{tabular} \label{tab:FeasStudyResult}
\end{table}

\begin{figure}
\centering
\subfigure[Fraction and accumulated dose for the \textit{non-adaptive robust stratetegy} combined with $\mathbb{E}$-optimization.]{\includegraphics[scale=0.25]{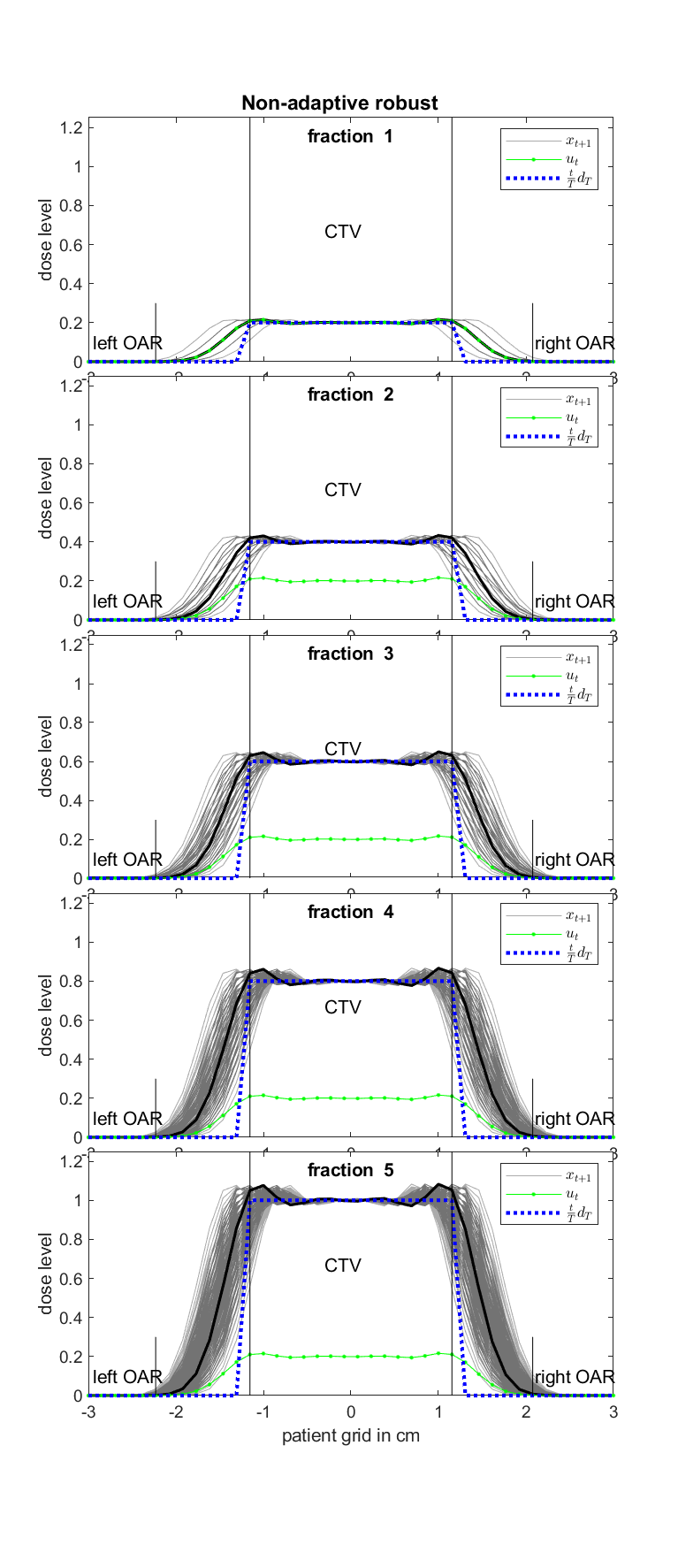} \label{fig:FractDose_Exp_Static_fn_5}}~ 
\subfigure[Fraction and accumulated dose for the \textit{time-varying adaptive strategy} combined with $\mathbb{E}$-optimization.]{\includegraphics[scale=0.25]{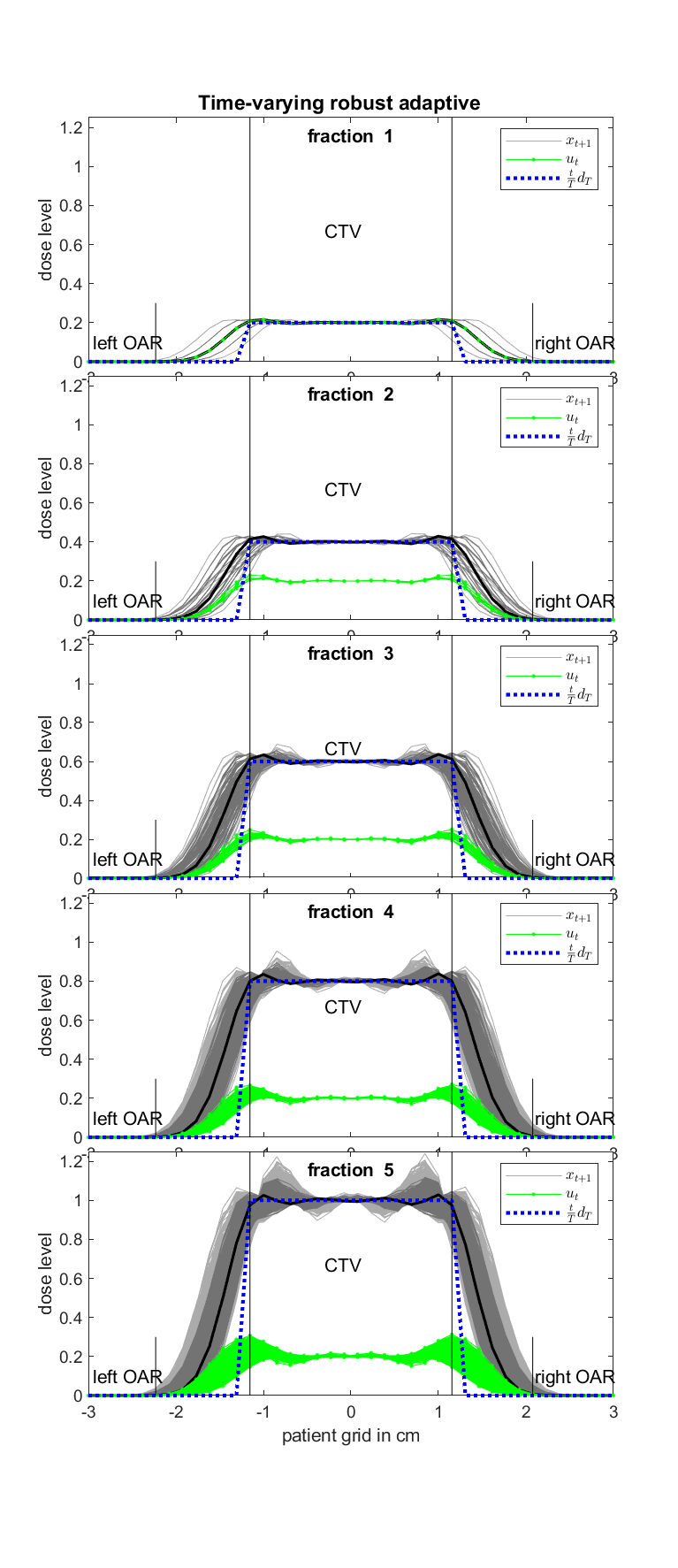} \label{fig:FractDose_Exp_RobART_fn_5}}
\subfigure[Fraction and accumulated dose for the \textit{time-varying adaptive strategy} combined with worst-case-optimization.]{\includegraphics[scale=0.25]{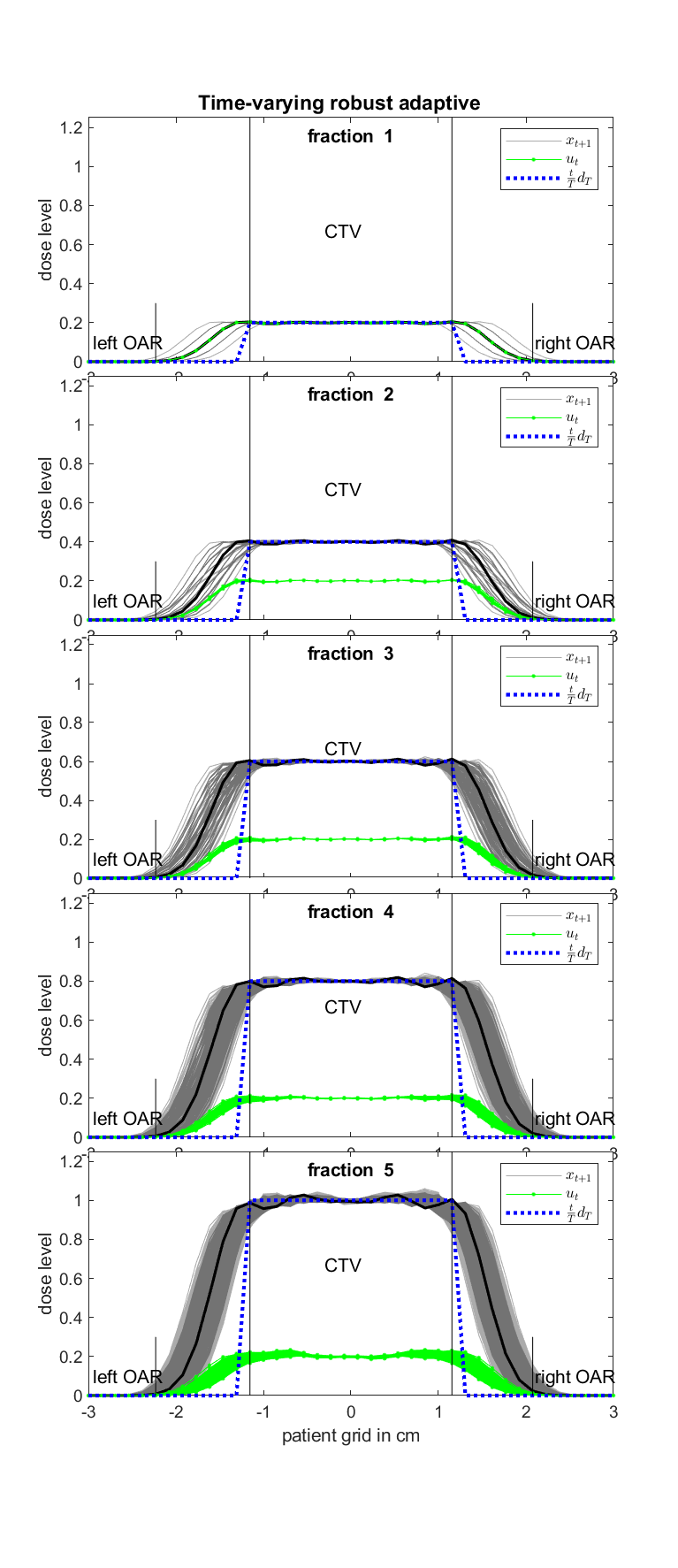} \label{fig:FractDose_Minimax_RobART_fn_5}}
\subfigure[Fraction and accumulated dose for the \textit{time-varying adaptive strategy} combined with CVaR-optimization.]{\includegraphics[scale=0.25]{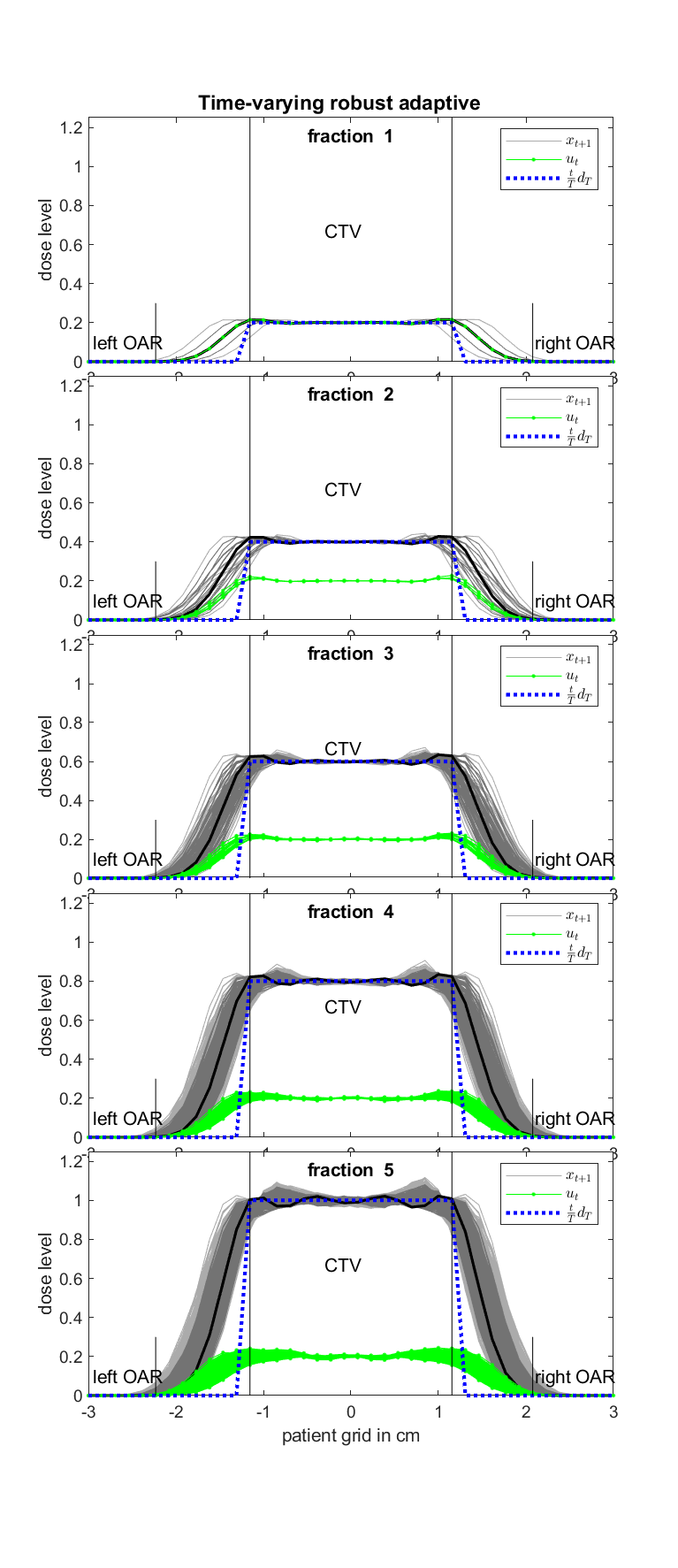} \label{fig:FractDose_CVaR_RobART_fn_5}}
\caption{Evaluation of the fraction doses and accumulated dose~$x_{t+1}$ for $t=0,\dots,T-1$ throughout the course of treatment with~$T=5$ fractions in comparison with the prescribed fraction dose~$t \frac{d_T}{T}$ for the stabilized \textit{non-adaptive robust} and \textit{time-varying adaptive strategy}. The most likely realization of~$x_{t+1}$ is represented by the solid black line, while the realizations within one standard deviation are visualized in dark grey.}
\label{fig:FractDose_Exp_WC_CVaR_fn_5}
\end{figure}

\begin{figure}
\centering
\subfigure[Fraction and accumulated dose for the unstable \textit{time-and-scenario-dependent robust adaptive strategy} combined with $\mathbb{E}$-optimization.]{\includegraphics[scale=0.4]{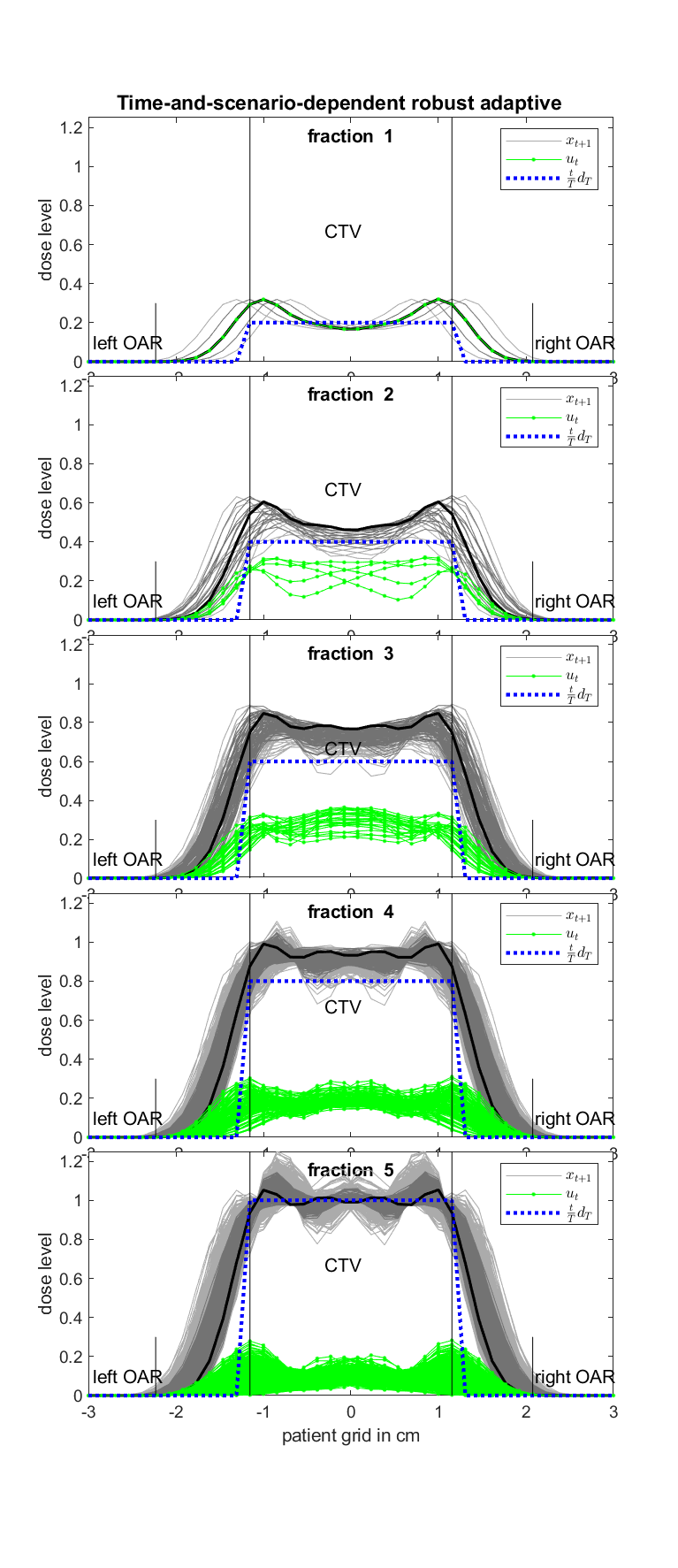} \label{fig:FractDose_Exp_nostab_Optimal_fn_5}}~ 
\subfigure[Fraction and accumulated dose for the unstable \textit{MPC-strategy} with~$K=3$ combined with $\mathbb{E}$-optimization.]{\includegraphics[scale=0.4]{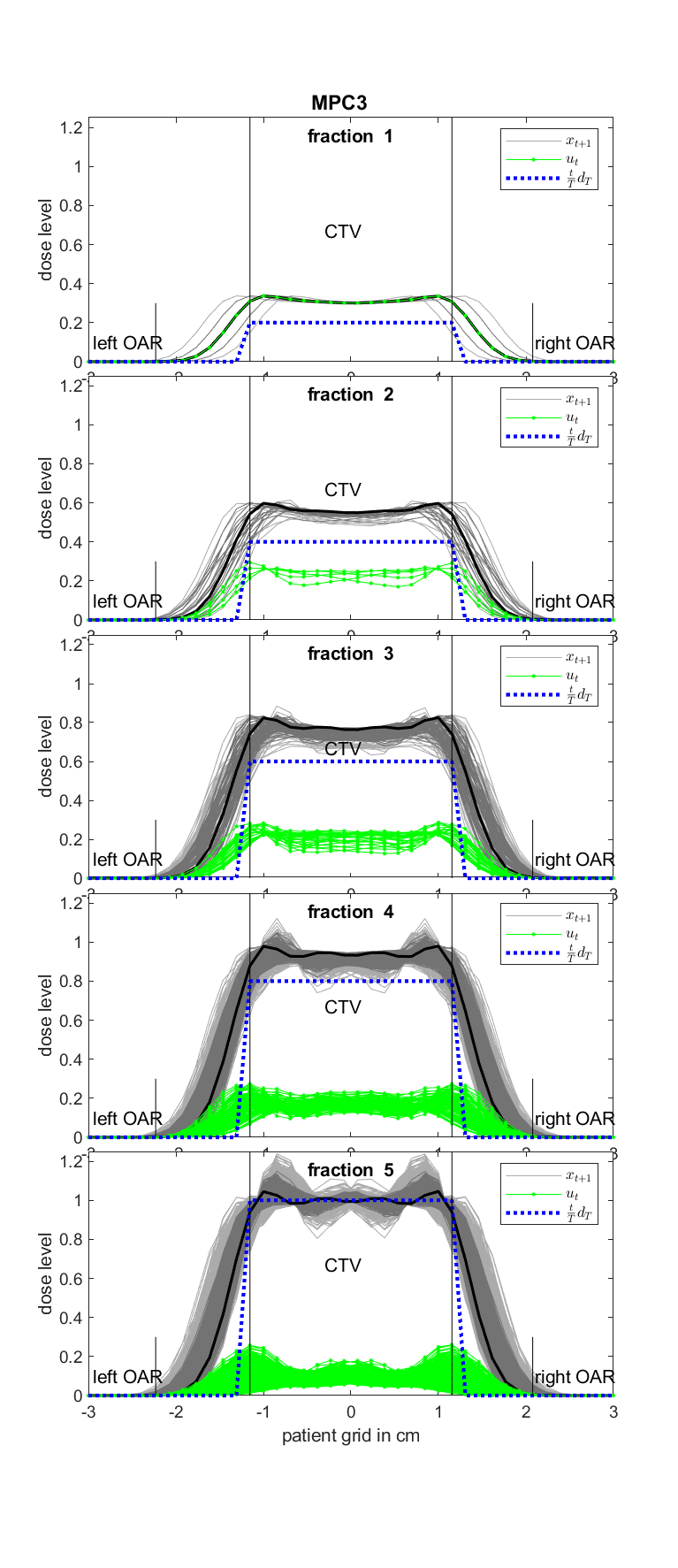} \label{fig:FractDose_Exp_nostab_MPC3_fn_5}}~
\subfigure[Fraction and accumulated dose for the unstable \textit{MPC-strategy} with~$K=3$ combined with worst-case-optimization. ]{\includegraphics[scale=0.4]{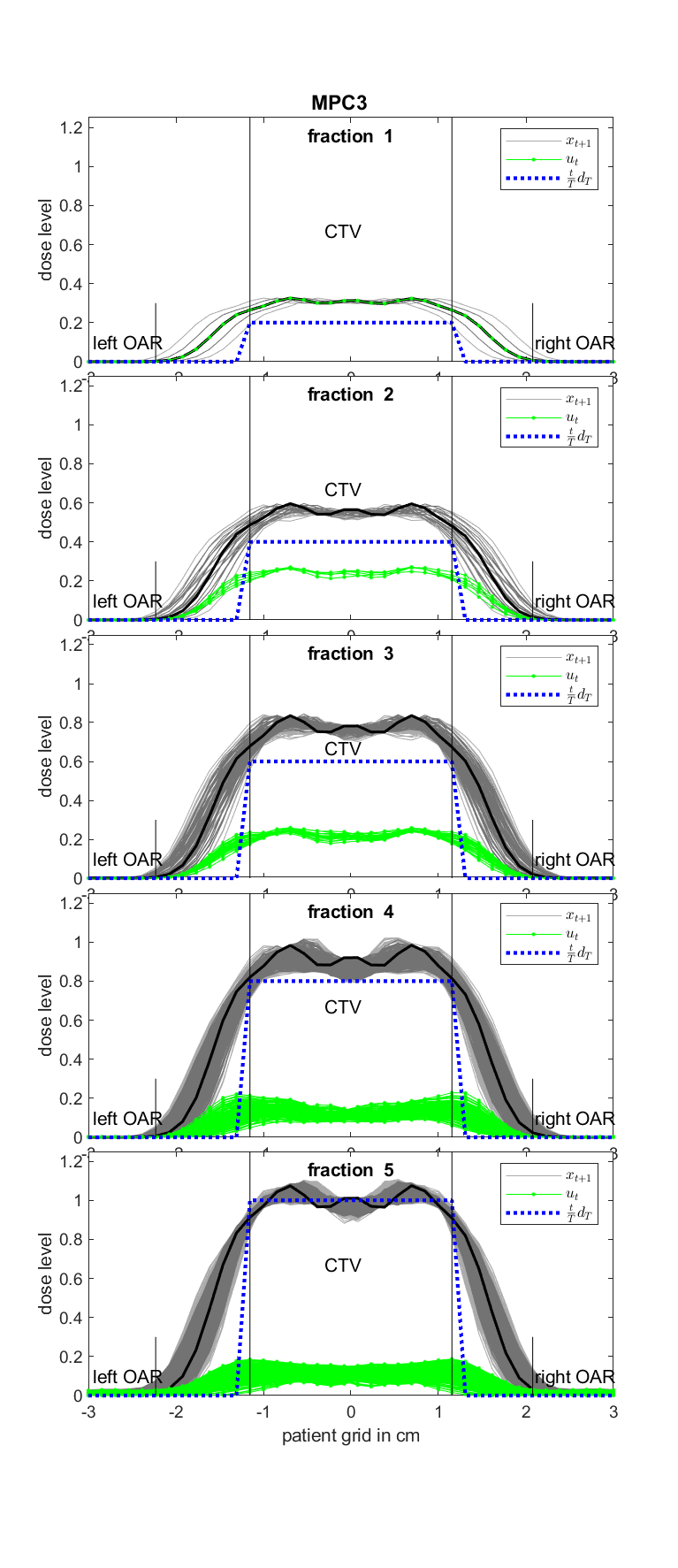} \label{fig:FractDose_Minimax_nostab_MPC3_fn_5}}
\caption{Evaluation of the fraction doses and accumulated dose~$x_{t+1}$ for $t=0,\dots,T-1$ throughout the course of treatment with~$T=5$ fractions in comparison with the prescribed fraction dose~$t \frac{d_T}{T}$ for the unstable \textit{time-and-scenario-dependent robust adaptive} and \textit{MPC-strategies} with~$K=3$. The most likely realization of~$x_{t+1}$ is represented by the solid black line, while the realizations within one standard deviation are visualized in dark grey.}
\label{fig:FractDose_Exp_nostab_fn_5}
\end{figure}

\begin{figure}
\centering
\subfigure[Fraction and accumulated dose for the stabilized \textit{time-and-scenario-dependent robust adaptive strategy} combined with $\mathbb{E}$-optimization.]{\includegraphics[scale=0.4]{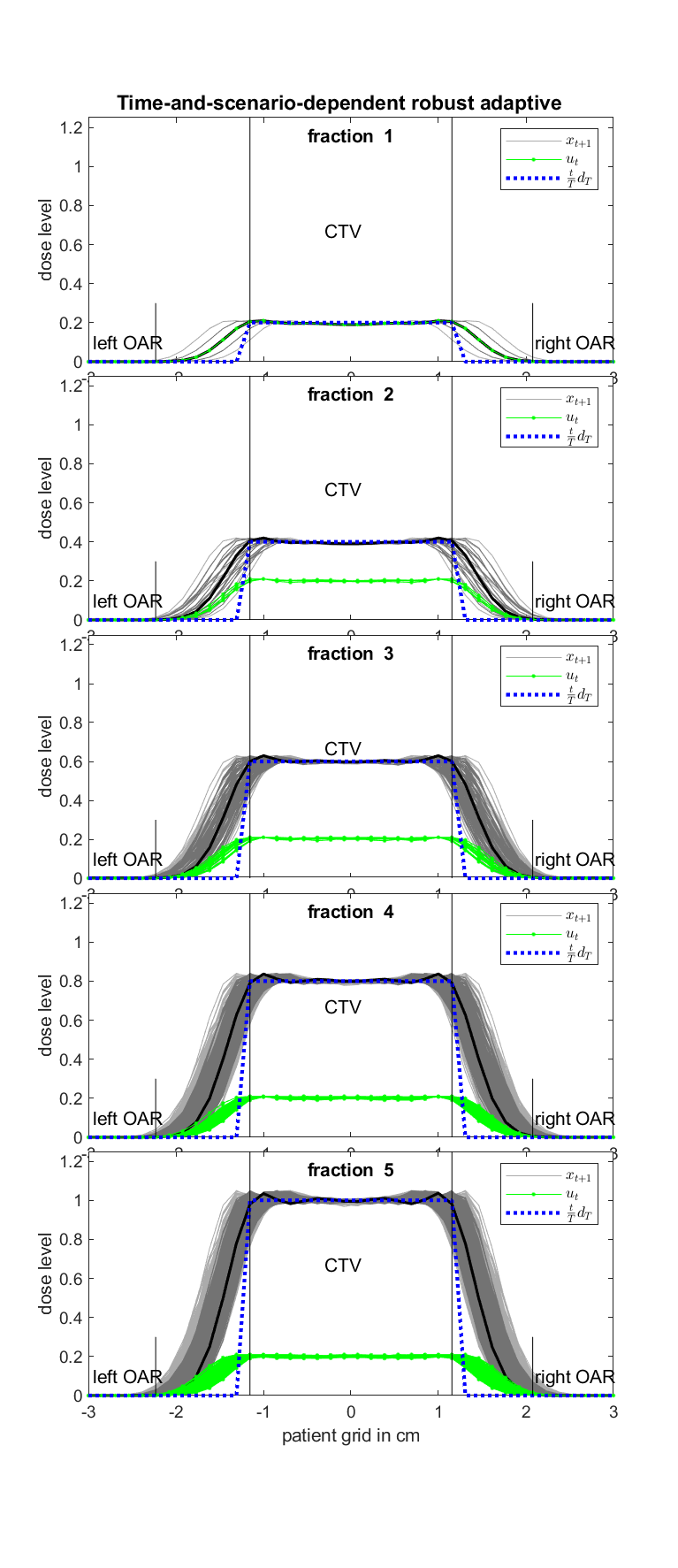} \label{fig:FractDose_Exp_stab_Optimal_fn_5}}~ 
\subfigure[Fraction and accumulated dose for the stabilized \textit{MPC-strategy} with~$K=3$ combined with $\mathbb{E}$-optimization.]{\includegraphics[scale=0.4]{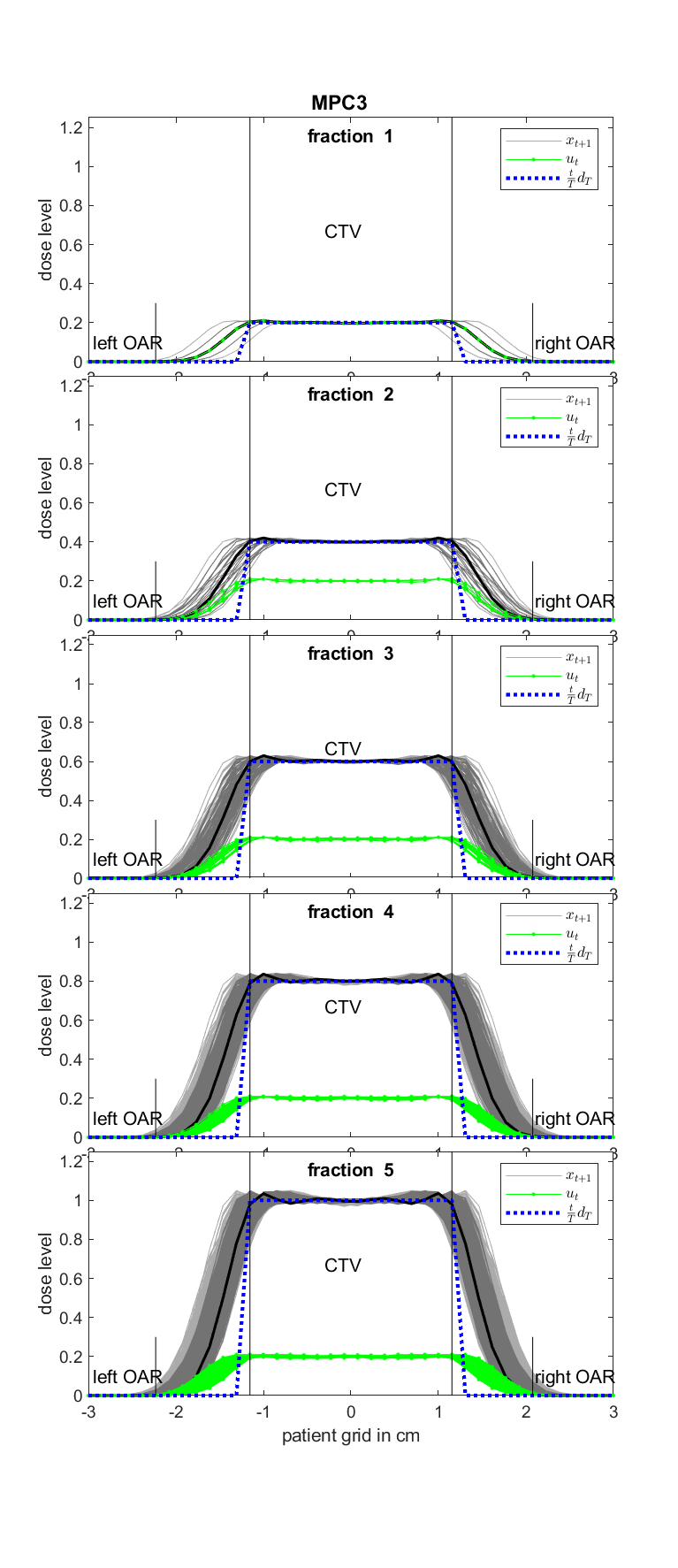} \label{fig:FractDose_Exp_stab_MPC3_fn_5}}~
\subfigure[Fraction and accumulated dose for the stabilized \textit{MPC-strategy} with~$K=3$ combined with worst-case-optimization. ]{\includegraphics[scale=0.4]{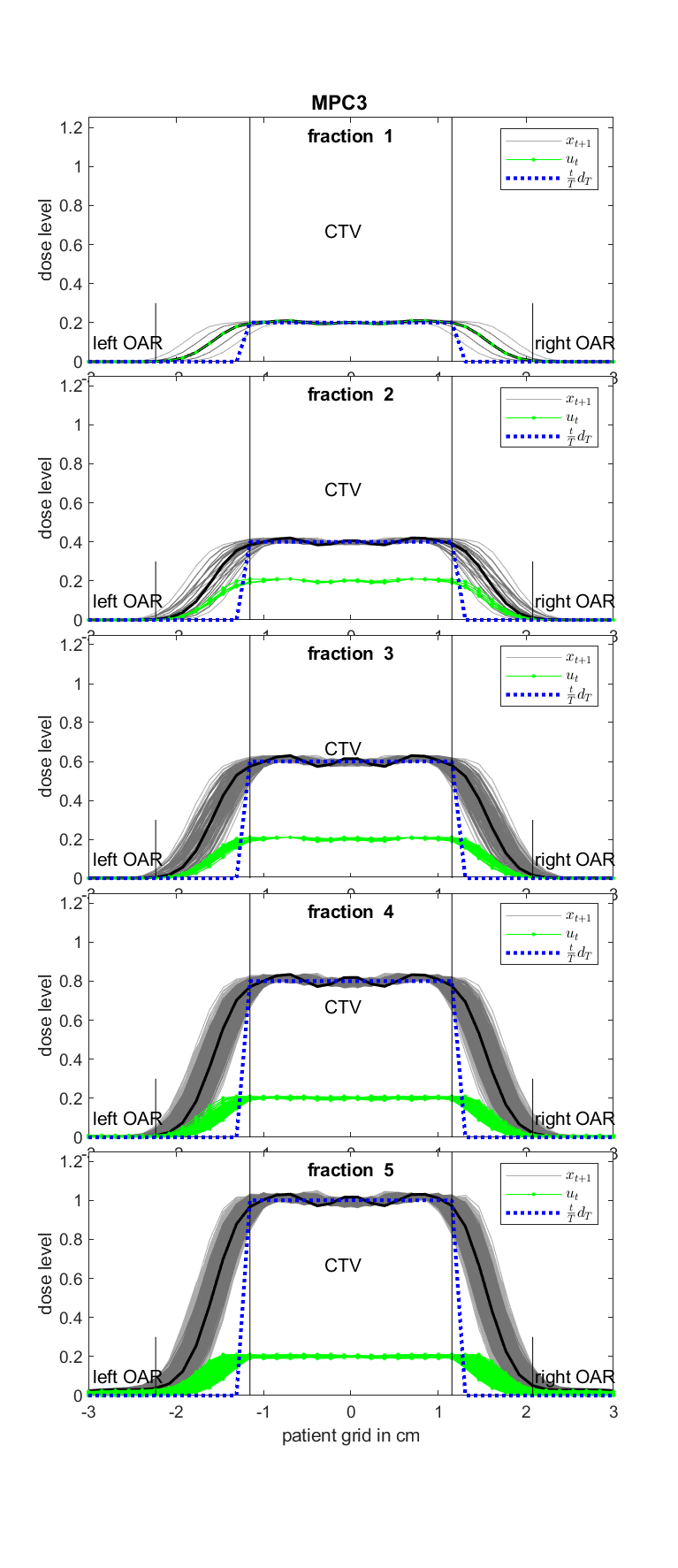} \label{fig:FractDose_Minimax_stab_MPC3_fn_5}}
\caption{Evaluation of the accumulated dose~$x_{t+1}$ and plans~$u_t$ for $t=0,\dots,T-1$ throughout the course of treatment with~$T=5$ fractions in comparison with the prescribed fraction dose~$t \frac{d_T}{T}$ for the stabilized \textit{time-and-scenario-dependent robust adaptive} and \textit{MPC-strategies} with~$K=3$. The most likely realization of~$x_{t+1}$ is represented by the solid black line, while the realizations within one standard deviation are visualized in dark grey.}
\label{fig:FractDose_Exp_stab_fn_5}
\end{figure}

In terms of evaluating the resulting treatment quality throughout a hypofractionated treatment, we gain the following insights.
First, the \textit{non-adaptive robust} and \textit{time-varying robust adaptive strategy} generate quite uniform dose profiles, as illustrated in Figures~\ref{fig:FractDose_Exp_Static_fn_5},~\ref{fig:FractDose_Exp_RobART_fn_5},~\ref{fig:FractDose_Minimax_RobART_fn_5} and~\ref{fig:FractDose_CVaR_RobART_fn_5}.
The x-axis represents the one-dimensional phantom where the location of the CTV and OARs is indicated through vertical separation lines.
The dose levels of the realized accumulated dose distributions and plans are given along the y-axis, ranging from 0 to 1.2 relative to the prescriptions dose level~$1$. 
The most likely accumulated dose realization in every fraction is indicated in black, while the realizations within one standard deviation are visualized in dark grey. The remaining less likely realizations are shown in light grey.
Second, the plans obtained from the \textit{time-varying robust adaptive strategy} in combination with expected-value-optimization become less uniform as the treatment progresses and the time horizon~$(T-t)$ shrinks, as demonstrated in Figure~\ref{fig:FractDose_Exp_RobART_fn_5}.  
These plans are optimized in an exact manner by taking advantage of the structure of the expected quadratic penalty, as demonstrated in Appendix~\ref{appendix:Reform_Expect}, which leads to a linear and quadratic dependency on the shrinking time horizon in the Hessian~$(T-t)\mathbb{E}\left[ S^T B^T D B S \right] + (T-t)(T-t-1)\mathbb{E}[S]^T B^T D B \mathbb{E}[S]$ of~(\ref{eq:Dquadcost}). 
Thus, as the plans for the last fractions are computed, the contribution of the squared mean value of~$(T-t)(T-t-1)\mathbb{E}[S]^T B^T D B \mathbb{E}[S]$ decreases and eventually disappears for~$t= T-1$. 
Therefore, the plans become less uniform because the term~$(T-t)\mathbb{E}\left[ S^T B^T D B S \right]$ squares the impact of the uncertainties modelled by the shift matrix~$S$.    
Since this is a result of minimizing the expected quadratic penalty, the dose profiles obtained from the \textit{time-varying adaptive strategy} combined with worst-case- and CVaR-optimization remain uniform, as illustrated in Figure~\ref{fig:FractDose_Minimax_RobART_fn_5} and~\ref{fig:FractDose_CVaR_RobART_fn_5}, respectively. 
CVaR-optimization is a compromise between expected-value- and worst-case-optimization in terms of conservativeness which is reflected in the shape of the resulting dose profiles. 
Third, the plans obtained from the \textit{time-and-scenario-dependent robust adaptive} and the \textit{MPC-strategies} are optimized in order to account for the possible realizations of uncertainty scenarios over the horizon~$T$ or~$K$, which may cause overcompensation as mentioned earlier.
The order of magnitude of overcompensation seems to correlate with the planning horizon~$T$ and~$K$. 
We observe that the longer the planning horizon, the larger the overcompensation will be at the edges of the CTV which are most likely to be affected by interfractional uncertainties, as illustrated in Figure~\ref{fig:FractDose_Exp_nostab_Optimal_fn_5} and~\ref{fig:FractDose_Exp_nostab_MPC3_fn_5} for the combination of expected value-optimization with \textit{time-and-scenario-dependent strategy} and the \textit{MPC-strategy} with~$K=3$, respectively. 
In expected value optimization, the overcompensation takes shape in the form of overdose at the edges of the CTV.
In worst-case-optimization, the impact of future uncertainties is handled by a flat dose-fall-off at the edges in order to reduce the risk to the nearby OARs in the event of the largest uncertainty, as shown in Figure~\ref{fig:FractDose_Minimax_nostab_MPC3_fn_5}.
Furthermore, worst-case-optimization results in a smaller spread of the various realized accumulated dose distributions~$x_{t+1}$. 
As a result of the stabilizing constraints, the accumulated doses and plans are more uniform, as illustrated in Figure~\ref{fig:FractDose_Exp_stab_Optimal_fn_5},~\ref{fig:FractDose_Exp_stab_MPC3_fn_5}, and~\ref{fig:FractDose_Minimax_stab_MPC3_fn_5}.
Moreover, the spread of dose realizations around the most likely outcome is decreased.
In the \textit{non-adaptive robust} and \textit{time-varying robust adaptive strategy}, the stabilizing constraints are not imposed since these strategies generate plans which do not exceed the prescribed fraction dose across the whole CTV. 
\subsection{Conventional fractionation length study}
In the \textit{conventional fractionation length study}, we evaluate the interplay between robustness and non-adaptive and adaptive strategies for a conventionally long treatment with~$30$ fractions.
The performance of the~\textit{non-adaptive robust},~\textit{time-varying robust adaptive},~\textit{MPC1-},~\textit{MPC2-} and~\textit{MPC3-strategy} combined with expected value-, worst-case- and CVaR-optimization is measured by evaluating their final dose distribution~$x_T$ for a population of~$100$ treatments with predictable interfractional uncertainties. 

\begin{figure}
\centering
\subfigure[Dose-probability-histogram for non-adaptive and adaptive strategies combined with $\mathbb{E}$ optimization.]{\includegraphics[scale=0.33]{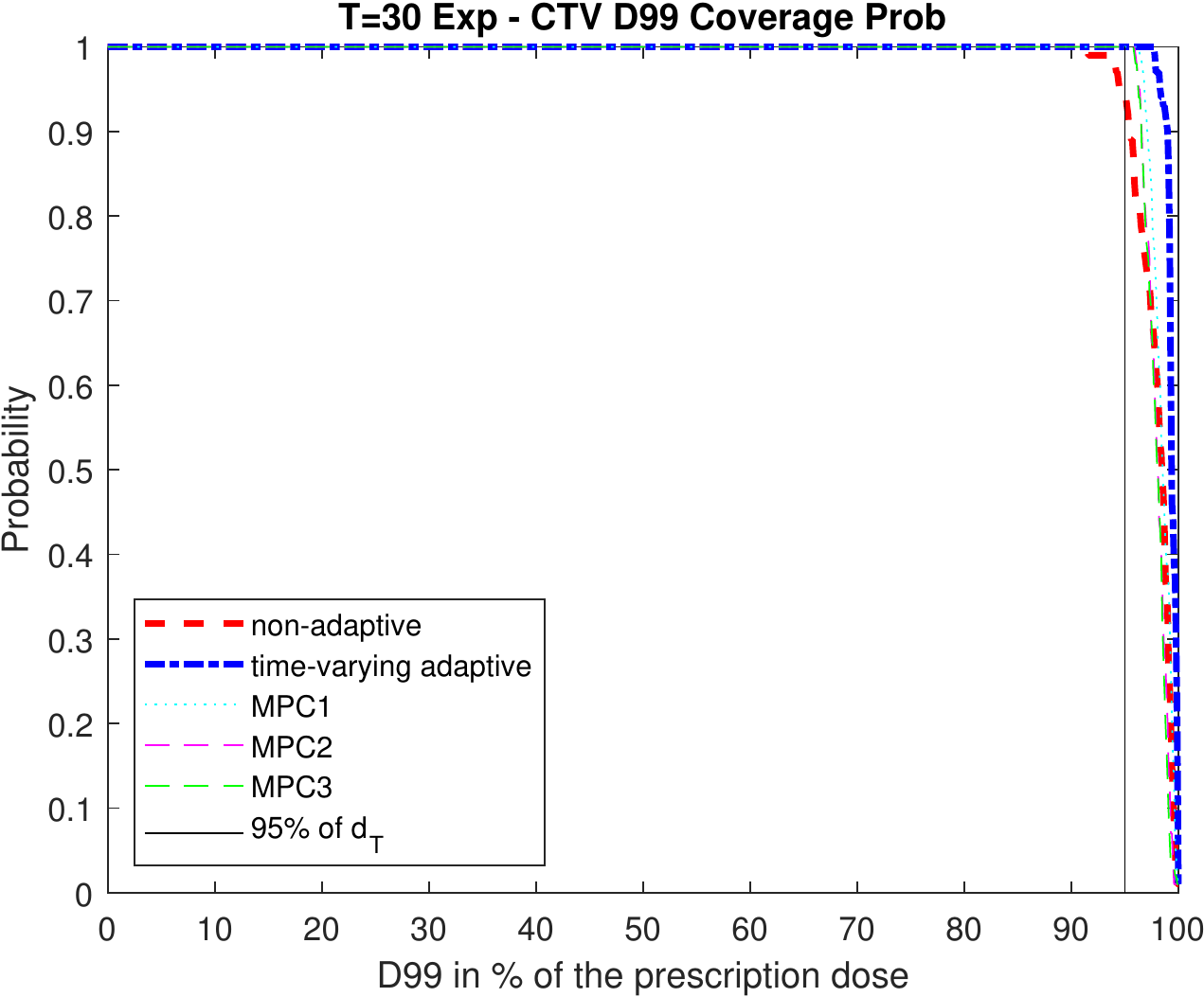} \label{fig:Exp_Pred_T30_CTVCovProb}}~ 
\subfigure[Dose-probability-histogram for non-adaptive and adaptive strategies combined with worst-case-optimization.]{\includegraphics[scale=0.33]{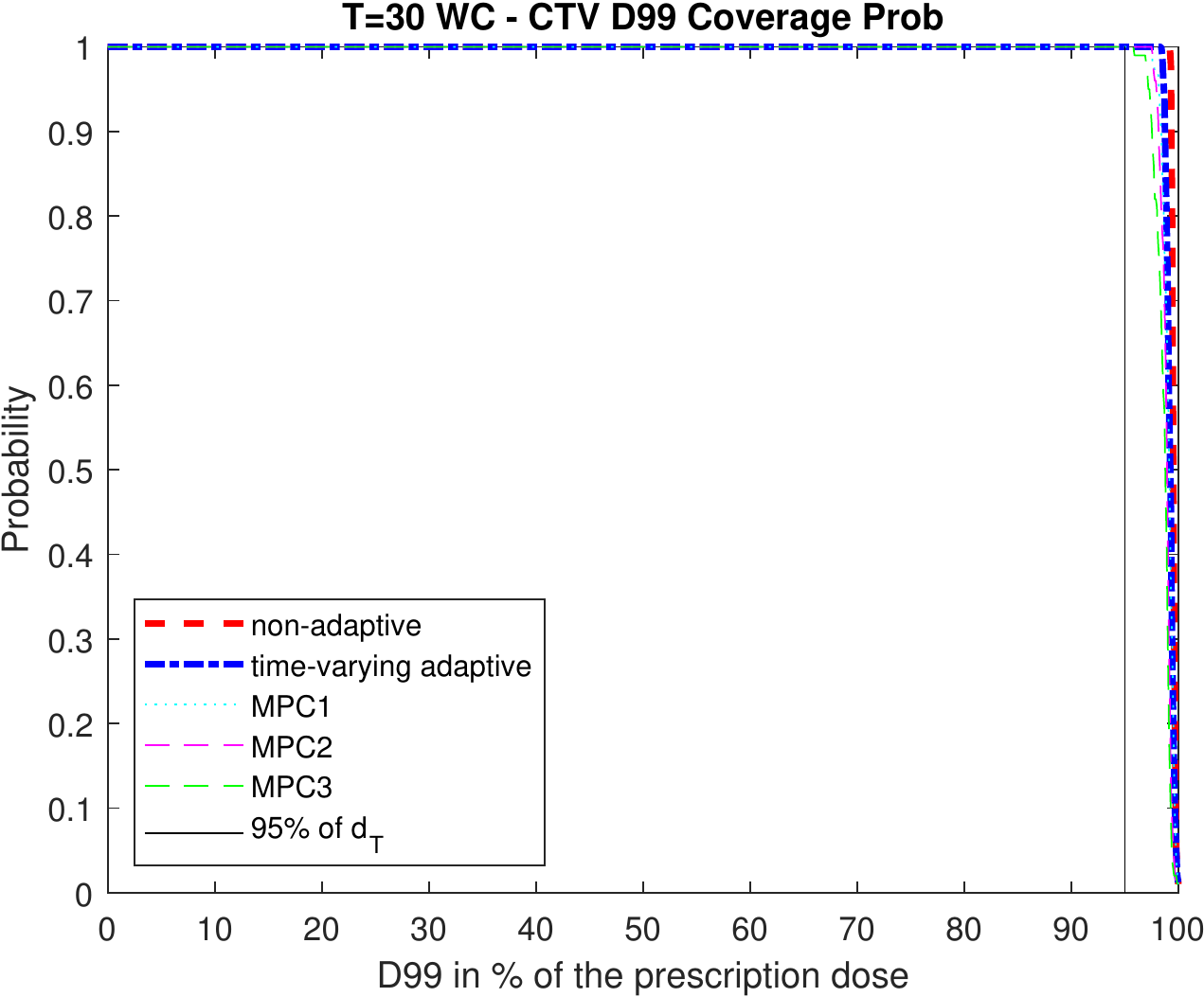} 	\label{fig:Minimax_Pred_T30_CTVCovProb}}~ 
\subfigure[Dose-probability-histogram for non-adaptive and adaptive strategies combined with CVaR optimization.]{\includegraphics[scale=0.33]{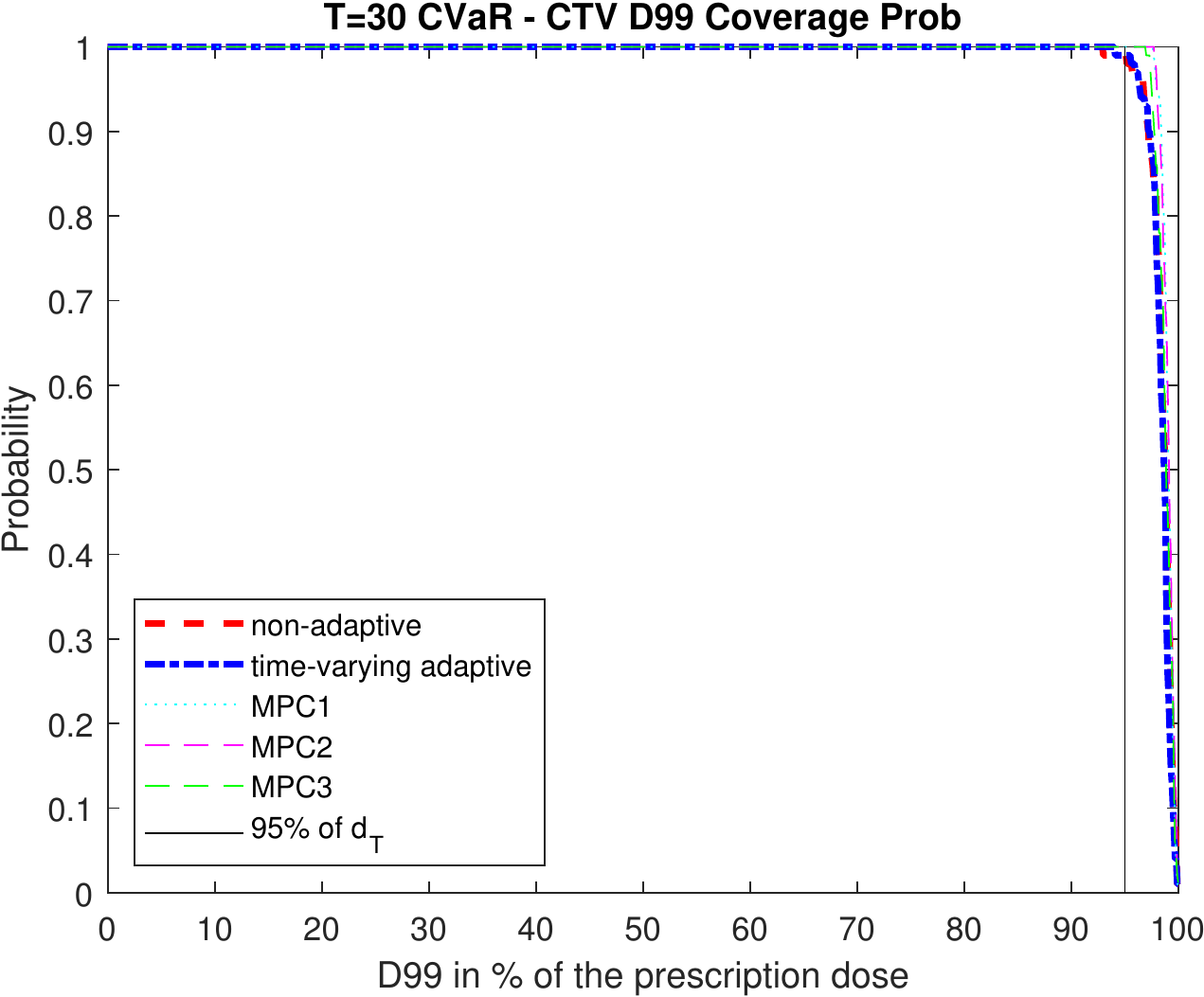} \label{fig:CVaR_Pred_T30_CTVCovProb}}~  
\caption{Evaluation of the final dose~$x_T$ in the CTV for the robust models and adaptive strategies by using dose-probability-histograms to compare the probability that $99\%$ of the CTV receives a certain percentage of the prescription dose or higher after exposure to predictable uncertainties.}
\label{fig:T30_CTV_CovProb}
\end{figure}

Their performance in terms of CTV-coverage under uncertainty is illustrated by dose-probability histograms in Figure~\ref{fig:T30_CTV_CovProb}. 
The dose-probability histograms show the probability that the final accumulated dose distribution~$x_T$ provide CTV-coverage in accordance with the ICRU guidelines~\cite{ICRU62} which recommend that 99\% of the CTV should receive at least 95\% of prescription dose. 
The probability that~$99\%$ of the CTV will receive at least a certain dose level or above is plotted on the y-axis. 
In order to facilitate the analysis of the performance of the various robust non-adaptive and adaptive strategies, the recommended dose level of~$95\%$ of the prescription dose is displayed in Figure~\ref{fig:T30_CTV_CovProb}. 
According to the results illustrated in Figures~\ref{fig:Exp_Pred_T30_CTVCovProb},~\ref{fig:Minimax_Pred_T30_CTVCovProb} and~\ref{fig:CVaR_Pred_T30_CTVCovProb}, the majority of strategies and robust models appear to provide sufficient target coverage as given by the ICRU-guidelines with similarly high probability.
This result indicates that robust optimization in combination with the conventional non-adaptive strategy ensures CTV-coverage, if the uncertainties occurring throughout the course of treatment are predicted correctly and are included in the robust optimization problem.
Among the investigated robust models, worst-case-optimization may be the most favourable model to be combined with the non-adaptive treatment approach, as  demonstrated in Figure~\ref{fig:Minimax_Pred_T30_CTVCovProb}.

\begin{figure}
\centering
\subfigure[Evaluation of the~$90^{th}$ percentile in the right OAR over 100 treamtents following the non-adaptive and adaptive-strategies combined with $\mathbb{E}$-optimization.]{\includegraphics[scale=0.4]{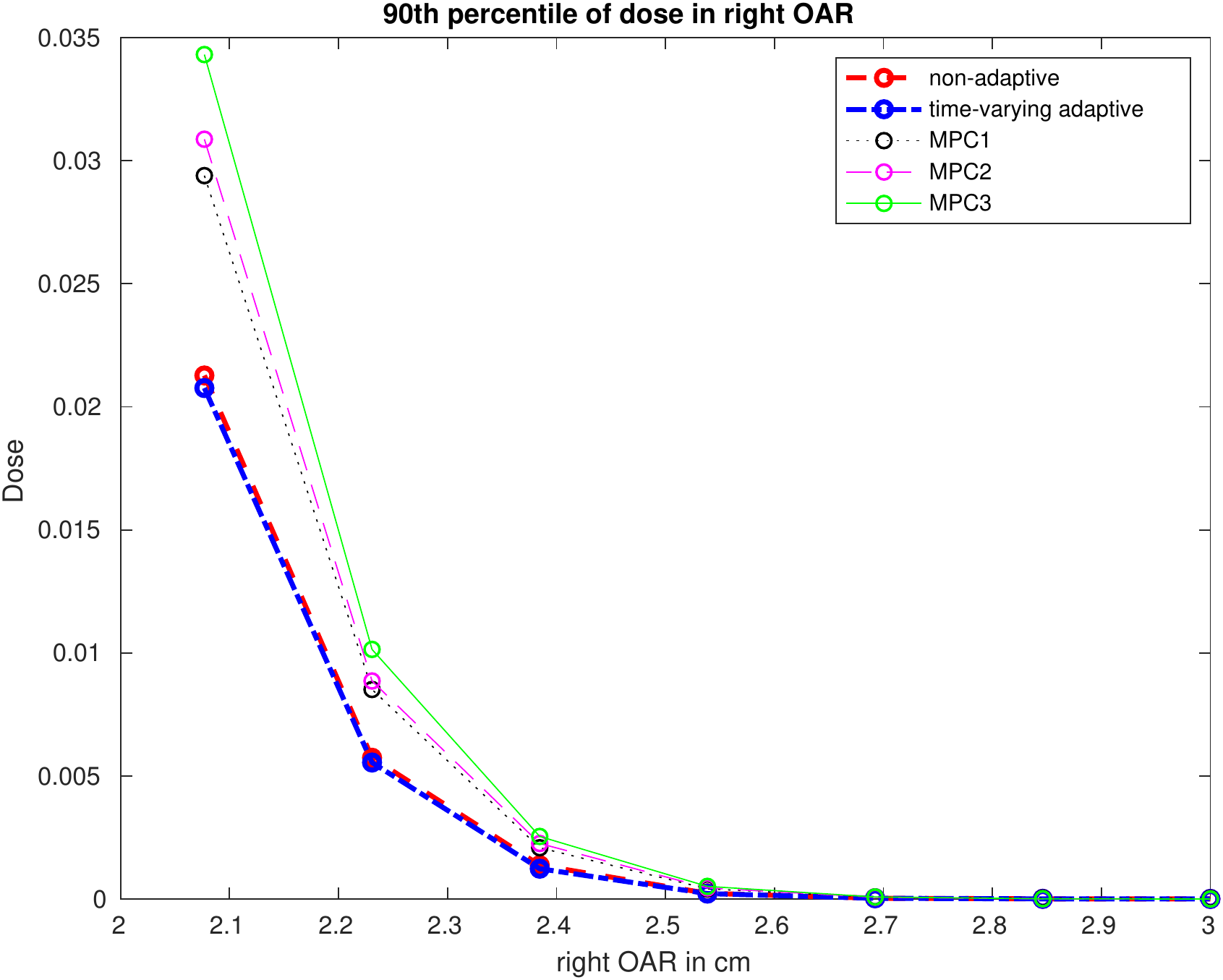} \label{fig:Exp_Pred_T30_rOAR_90thPercentile}}~ 
\subfigure[Evaluation of the~$90^{th}$ percentile in the right OAR over 100 treamtents following the non-adaptive and adaptive-strategies combined with worst-case-optimization.]{\includegraphics[scale=0.4]{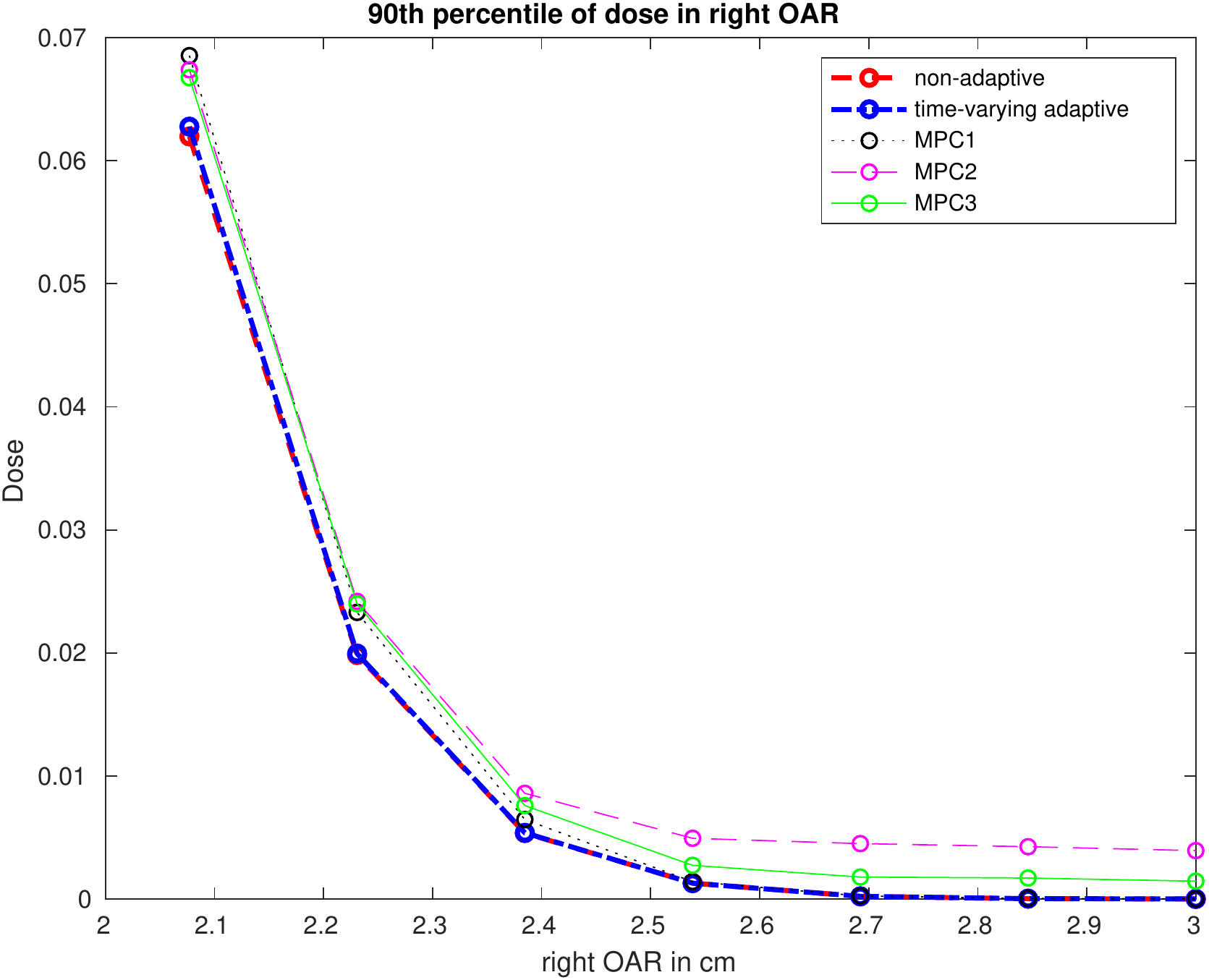} \label{fig:Minimax_Pred_T30_rOAR_90thPercentile}}
\caption{Evaluation of the~$90^{th}$ percentile of the final dose~$x_T$ in the right OAR over all simulated treatments exposed to interfractional uncertainties. during treatments following the non-adaptive and adaptive strategies combined with expected value- and worst-case-optimization model.}
\label{fig:rOAR_90thPercentile}
\end{figure}
Concerning the interplay between robust models and the adaptive strategies, expected value optimization seems to be a good match for the \textit{time-varying robust adaptive strategy}, which achieves target coverage with the highest probability compared to the \textit{MPC-strategies}, as shown in Figure~\ref{fig:Exp_Pred_T30_CTVCovProb}. 
According to our evaluation of CTV-coverage probability for the various robust strategies combined with expected value optimization in Figure~\ref{fig:Exp_Pred_T30_CTVCovProb}, the \textit{MPC-strategies} which are based on time-and-scenario optimization variables may not be as a good match for expected value optimization as the \textit{time-varying robust adaptive strategy}.
However, it should be noted that the \textit{MPC-strategies} are approximative methods in contrast to the \textit{time-varying adaptive strategy}, which in combination with expected value optimization gives exact solutions.  
Since the expected value over the anticipated scenarios gives little weight to extreme scenarios with a low probability, the use of time-and-scenario-dependent optimization variables may not be suitable.
On the other hand, time-and-scenario-dependent optimization variables appear to be good combination for the more conservative optimization models as shown in Figure~\ref{fig:Minimax_Pred_T30_CTVCovProb} and~\ref{fig:CVaR_Pred_T30_CTVCovProb} according to which the \textit{MPC-strategies} provide target coverage with the highest probability.
In the linear formulations of our worst-case- and CVaR-optimization problems~(\ref{eq:dualCVaR}) and~(\ref{eq:LinMinimax}), every realization of uncertainty scenarios is represented by a corresponding constraint. 
Time-and-scenario-dependent optimization variables seem to exploit this feature and thus, the additional computational effort may be worthwhile in the case of combining MPC-strategies with conservative robust optimization models.  
Based on the observation that all \textit{MPC-strategies} show very similar trends in terms of CTV-coverage, the most suitable prediction horizon cannot be clearly identified. 
The similarity of these trends among the MPC-strategies may be caused by their rather short prediction horizons relative to the treatment length of~$30$ fractions.

Concerning the risk of high dose exposure in the OAR, the accumulated dose in the OARs is evaluated by evaluating the dose levels of the~$90^{th}$ percentile over the 100 treatments.
We evaluate the~$90^{th}$ percentile in order to visualize the order of magnitude of an upper dose threshold for~$90\%$ of the treatments with the robust optimization models which according to our framework are most likely to provide sufficient CTV-coverage.
Since the right OAR is closer to the CTV, it may likely receive a higher dose than the left OAR in the event of interfractional uncertainties. 
Since expected value- and worst-case-optimization appear to provide the highest probability of CTV-coverage, we choose to draw attention to their impact on the right OAR.
The voxels of the right OAR which are in proximity to the CTV receive the largest amount of dose as shown in Figure~\ref{fig:Exp_Pred_T30_rOAR_90thPercentile} and~\ref{fig:Minimax_Pred_T30_rOAR_90thPercentile}, which illustrate the ~$90^{th}$ percentile for each voxel in the right OAR after 100 treatments.
In case of expected value optimization, the \textit{time-varying robust adaptive strategy} leads to the least dose exposure to the right OAR among the adaptive strategies, but its dose levels are comparable to those of non-adaptive strategy. 
Since the combination of the \textit{time-varying robust adaptive strategy} with expected value optimization gives the highest probability of CTV-coverage, we may conclude that a robust adaptive strategy based on expected value optimization may be most effective with time-and-scenario-independent optimization variables. 
In case of worst-case-optimization, all strategies lead to similar upper dose levels in the right OAR, indicating that robust adaptive strategies may not increase the risk of high dose exposure to the OARs.  
Moreover, the upper dose thresholds of the accumulated dose in the right OAR are higher than in combination with expected value optimization.
However, these differences between non-adaptive and adaptive strategies and the robust models are rather small since the simulated treatments follow the same distribution as those included in the robust models.

Overall, we conclude that if the treatment is supposed to be adaptive with a high adaptation frequency throughout the course of treatment , it may be the most beneficial to combine expected value optimization with time-and-scenario-independent optimization variables.
If however, the treatment is supposed to be executed in a non-adaptive manner or only adaptive in rare cases, a more conservative model in combination with time-and-scenario-dependent optimization variables may be recommended.

\section{Discussion}
In this study, a proof-of-concept framework is presented which handles interfractional variations by employing various robust adaptive strategies based on either time-and-scenario-independent, time-dependent or time-and-scenario-dependent optimization variables which are combined with expected value-, worst-case- or CVaR-optimization.  
The various strategies are evaluated and compared with each other based on their performance to mitigate the impact of uncertainties on the accumulated dose and their ability to steer the accumulated dose as close as possible to the prescribed dose distribution. 
In particular, we evaluate the potential benefits of increasing the degrees of freedom in the optimization variables and their interplay with robust optimization models with varying grades of conservativeness.
Furthermore, we investigate the performance of our framework in a hypofractionated and conventionally long radiotherapy treatment.
In order to conduct the comparative analysis of the various strategies in a fair manner, all strategies are exposed to the same set of interfractional uncertainty scenarios. 
In the analysis of our framework in the hypofractionated treatment, the objective function values of the robust adaptive strategies at their optimal solutions are computed and as a consequence, conclusions are drawn based on objective criteria. 
In the analysis of our framework in the conventionally long treatment,
the final accumulated dose distributions for the various robust non-adaptive and adaptive strategies are evaluated in terms of CTV-coverage probability and high dose exposure to the right OAR after 100 simulated treatments in the presence of interfractional uncertainties.
Overall, we gain the following insights.
First, we learn from the analysis of the \textit{non-adaptive} and \textit{time-varying robust strategy}, that a time-dependent problem can be solved in an optimal manner by solving the corresponding problem with time-independent optimization variables. 
Thus, we can implement an adaptive treatment strategy by dynamically solving a sequence of time-independent problems as is the case in the \textit{time-varying adaptive strategy}. 
This finding is especially relevant in a clinical context since it implies that adaptive radiation therapy can become part of clinical practice using  commercially available treatment planning software. 
Second, we learn that time-and-scenario-dependent optimization variables, used in the \textit{time-and-scenario-dependent robust adaptive} and \textit{MPC-strategies} in order to account for dose predictions, are most compatible with conservative robust optimization models and that including dose predictions potentially improves treatment quality compared to the non-adaptive treatment approach.
Third, we learn if the uncertainties occurring during treatment coincide with those included in the robust optimization models, our non-adaptive robust strategies give high probability of sufficient CTV-coverage without putting the OARs at greater risk.
This finding implies that the robust non-adaptive approach leads to a successful treatment for that group of patients whose interfractional uncertainties have been predicted correctly.
Among the evaluated robust models, worst-case-optimization may be the most suitable match for the non-adaptive strategy.
Fourth, we learn that an adaptive strategy intended for daily adaptation does not require a very conservative robust optimization model in order to provide sufficient CTV-coverage. 
In case the robust model is based on expected-value-optimization, our study suggests that time-and-scenario-independent optimization variables may be most compatible.

Throughout this work, the emphasis is put on the mathematical properties of the proposed strategies and understanding the role of robust optimization in adaptive radiation therapy.
The insights gained from the mathematical evaluation and Proposition~\ref{prop1} are geometry-independent, since the system dynamics and robust optimization problems can be applied to two- or three-dimensional geometries.
In the computational study however, we consider a one-dimensional patient phantom~\cite{Boeck2017,Lof1998} and subjected it to interfractional variations which are modeled as rigid whole-body shifts~\cite{VanHerk2000,Albin2012,Boeck2017} in order to reduce computational effort. 
The assumption of having i.i.d.~shifts is reasonable for random setup errors and organ motion~\cite{VanHerk2000,Unkelbach2009,Unkelbach2004,Albin2012}.
Moreover, the evaluation of our proposed framework in an idealized model gives the opportunity to solely focus on their mathematical properties and to better understand the interplay between robustness and the various non-adaptive and adaptive strategies. 
Gaining insight into the role of the problem formulation is critical before applying our framework to clinical patient data. 
Evaluating our framework first on clinical patient data would make it more challenging to identify the generic properties of the proposed framework, since they could potentially coincide with a specific anatomy or beam setup.
Since our framework is evaluated on a simplified geometry, the differences between the various strategies might be more or less pronounced for a two- or three-dimensional geometry.
As a consequence of assuming i.i.d.~interfractional uncertainties, complete information about the accumulated dose and no costs associated with plan adaptation, the results of our study should be interpreted in relative terms.  

Overall, this study provides mathematical insights into the requirements and suitable set up for a robust adaptive framework necessary to further develop our framework to include more complex model assumptions such as organ deformation and adapting the robust optimization model in response to unpredictable uncertainty scenarios.
To the best of our knowledge, this is the first study of its kind to address the mathematical properties required for robust adaptive radiation therapy. 
Thus, this work represents a fundamental study which gives the opportunity to further study robust adaptive radiation therapy planning.    

\section{Conclusion}
In this paper, we introduce a dynamic planning framework to handle interfractional geometric variations in which robust adaptive strategies with varying degrees of freedom and grade of conservativeness are presented and evaluated.
The motivation behind introducing and analysing such a wide range of robust adaptive strategies is to better understand how to generate the right plan for every fraction in the event of interfractional uncertainties in order to mitigate their impact on the accumulated dose distribution throughout the course of treatment.
Therefore, a mathematical and computational study is carried out in order to study the implications of combining robust optimization models with adaptive radiation therapy strategies in the context of a hypofractionated and conventionally long radiotherapy treatment.

From our study we gain insights on the strengths of using time-and-scenario-independent optimization variables in robust adaptive planning and in combination with robust non-adaptive treatment strategies.
Under the model assumptions of optimizing the final accumulated dose in the presence of i.i.d.~interfractional uncertainties, the optimal plan for a time-varying planning approach can be obtained by dynamically solving the corresponding problem with time-and-scenario-independent optimization variables.
In non-adaptive robust planning, the achieved treatment quality is comparable to that of the robust adaptive strategies, if the actually occurring interfractional uncertainties coincide with those accounted for in the robust plan.
In terms of the interplay between robustness and adaptive planning, we learn that time-and-scenario-dependent optimization variables may be most compatible with worst-case-optimization in order to benefit the most from dose predictions in the adaptive plans, while expected value optimization does not benefit from time-and-scenario-dependent optimization variables to the same extent.

The clinical value of our proposed framework is the insight into the strengths and weaknesses of the conventional non-adaptive approach in combination with robust optimization models, and in which cases and ways robust adaptive strategies can be used to further improve treatment outcome.   
Overall, based on the findings of this work we have a better understanding about the requirements on the degrees of freedom of the optimization variables and their interplay with robust optimization models, which is crucial for further development of planning frameworks for robust adaptive radiation therapy in order to manage various types of interfractional uncertainties experienced by patients.

\section*{Acknowledgements}
The authors thank Johan Karlsson for the helpful discussion on MPC.

\section*{References}
\bibliographystyle{unsrt} 
\bibliography{References_MPC_Project}

\appendix

\section{Reformulation of the expected quadratic penalty}
\label{appendix:Reform_Expect}
In the case of minimizing the expected value of~$\left\Vert x_0 + \sum_{t=0}^{T-1}BS(\omega_{t+1})u  - d_T \right\Vert^2_D $ can be formulated as follows
\begin{multline*}
\mathbb{E}[\sum_{t=0}^{T-1} \left\Vert x_t + BS(\omega_{t+1})u -d_T \right\Vert_D^2] = \\
\sum_{t_1 = 1}^{|\Omega|} p_{t_1} \sum_{t_2 = 1}^{|\Omega|} p_{t_2}\cdots \sum_{t_T = 1}^{|\Omega|} p_{t_T} 
\left( x_0 + BS(\omega_{t_1})u + BS(\omega_{t_2})u + \cdots + BS(\omega_{t_T})u -d_T \right)^T\\
D\left( x_0 + BS(\omega_{t_1})u + BS(\omega_{t_2})u + \cdots + BS(\omega_{t_2})u -d_T\right)= \\ 
\sum_{t_1 = 1}^{|\Omega|} p_{t_1} \sum_{t_2 = 1}^{|\Omega|} p_{t_2}\cdots \sum_{t_T = 1}^{|\Omega|} p_{t_T} \left[u^TS^T(\omega_{t_1})B^TDBS(\omega_{t_1})u + u^TS^T(\omega_{t_2})B^TDBS(\omega_{t_2})u + \cdots \right.\\ + \left. u^TS^T(\omega_{t_T})B^TDBS(\omega_{t_T})u + u^TS^T(\omega_{t_1})B^TDB \left( S(\omega_{t_2}) + \cdots + S(\omega_{t_T}) \right)u \right. \\ 
\left. + u^TS^T(\omega_{t_2})B^TDB \left( S(\omega_{t_1}) + S(\omega_{t_3}) + \dots + S(\omega_{t_T}) \right)u  + \cdots \right. \\
\left. + u^TS^T(\omega_{t_T})B^TDB \left( S(\omega_{t_1}) + S(\omega_{t_2}) + \dots + S(\omega_{t_{T-1}}) \right)u \right.\\
  \left. + 2(x_0 -d_T)^T D \left( BS(\omega_{t_1}) + BS(\omega_{t_2})+ \cdots + BS(\omega_{t_T})\right)u +(x_0-d_T)^TD(x_0-d_T) \right]  
\end{multline*}
According to our model assumptions of i.i.d interfractional uncertainties ~$\omega \in \Omega$,
\begin{itemize}
\item $\sum_{t_i = 1}^{|\Omega|} p_{t_i} S(\omega_{t_i}) = \sum_{t_j = 1}^{|\Omega|} p_{t_j} S(\omega_{t_j}) = \mathbb{E}[S]\ \forall i,j = 1,2,\dots,T $ where~$i\neq j$,
\item  $\sum_{t_i = 1}^{|\Omega|} p_{t_i} S(\omega_{t_i})^T B^T D B S(\omega_{t_i})  = \sum_{t_j = 1}^{|\Omega|} p_{t_j} S(\omega_{t_j})^T B^T D BS(\omega_{t_j}) = \mathbb{E}[S^TB^TDBS] \ \forall i,j = 1,2,\dots,T $ where~$i\neq j$, and
\item $\sum_{t_i = 1}^{|\Omega|} p_{t_i} S(\omega_{t_i})^T B^TD B \sum_{t_j = 1}^{|\Omega|} p_{t_j}S(\omega_{t_j})^T = \sum_{t_j = 1}^{|\Omega|} p_{t_j} S(\omega_{t_j})^T B^T DB \sum_{t_i = 1}^{|\Omega|} p_{t_i}S(\omega_{t_i})^T = \mathbb{E}[S]^T B^T D B \mathbb{E}[S] \ \forall i,j = 1,2,\dots,T $ where~$i\neq j$,
\end{itemize}
the expected quadratic penalty can be written as the following quadratic optimization problem
\begin{equation}
u^T \left( T \mathbb{E}[S^T B^T DB S] + T(T-1) \mathbb{E}[S]^T B^T D B \mathbb{E}[S] \right) u  + 2 T (x_0 -d_T)^T D B  \mathbb{E}[S] u + (x_0-d_T)^TD(x_0-d_T).
\end{equation}
\section{Proof of Proposition 1}
\label{appendix:Proof}
\begin{proof}
Since~$f$ is permutation invariant, i.e., under a permutation~$\pi: [U]\rightarrow[U]$, the value of the function~$f$ remains unchanged~$f \left( [U]^{\pi}\right) = f([U])$.
Moreover,~$f$ is convex and by using Jensen's inequality for convex functions of~$x_1,x_2,\dots,x_n$ with weights~$\lambda_i \geq 0$,~$i = 1,\dots,n$,~$\sum_{i=1}^n \lambda_i = 1$, 
\[
f \left( \sum_{i=1}^n \lambda_i x_i \right) \leq \sum_{i=1}^n \lambda_i f(x_i),
\]
the convex combination of all possible permutations~$\pi: [U] \rightarrow [U]$ of the T-tuple~$[U]$ gives
\[
f \left( \frac{1}{T!} \sum_{k=1}^{T!} [U]^{\pi_k}\right) \leq 
\frac{1}{T!} \sum_{k=1}^{T!} f([U]^{\pi_k}).
\] 
Since~$f$ is permutation invariant, we get for the right-hand side that
\[\sum_{k=1}^{T!} f([U]^{\pi_k}) =  T!f([U]). \]
The argument of~$f$ on left-hand side gives the mean over all elements in the T-tuple~$[U]$ for every combination~$k$,
\[
\frac{1}{T!}\sum_{k=1}^{T!} [U]^{\pi_k}  = \frac{1}{T!} \sum_{k=1}^{T!}  \left(  u_0,u_1,\dots,u_{T-1} \right)^{\pi_k} = \frac{(T-1)!}{T!} \left( \sum_{k=0}^{T-1} u_k,\sum_{k=0}^{T-1} u_k,\dots,\sum_{k=0}^{T-1} u_k \right).
\]
Thus, the left-hand side is a T-tuple of the mean over all vectors~$u_k$ for~$k=0,1,\dots,T-1$,
\[f \left( \frac{1}{T!} \sum_{k=1}^{T!} [U]^{\pi_k}\right) = f\left( \frac{1}{T} \left( \sum_{k=0}^{T-1} u_k,\sum_{k=0}^{T-1} u_k,\dots,\sum_{k=0}^{T-1} u_k \right) \right) = f \left( \left[ \bar{U} \right] \right).
\]
Combining the reformulated forms of the right-hand and left-hand side we have that
\[f \left( \left[ \bar{U} \right] \right) \leq f \left( \left[ U \right] \right), \]
which proves the proposition.
\end{proof}

\end{document}